\documentclass[%
 reprint,
nofootinbib,
 amsmath,amssymb,
 aps,
]{revtex4-2}

\usepackage{graphicx}
\usepackage{dcolumn}
\usepackage{bm}

\begin{document}

\preprint{APS/123-QED}

\title{On the Inherent Dose-Reduction Potential of Classical Ghost Imaging}

\author{Andrew M. Kingston}
 \email{andrew.kingston@anu.edu.au}
 \altaffiliation[Also at ]{CTLab: National Laboratory for Micro Computed-Tomography, Advanced Imaging Precinct, The Australian National University, Canberra, ACT 2601, Australia}
 \affiliation{Department of Applied Mathematics, Research School of Physics, The Australian National University, Canberra, ACT 2601, Australia}

\author{Wilfred K. Fullagar}
 \affiliation{Department of Applied Mathematics, Research School of Physics, The Australian National University, Canberra, ACT 2601, Australia}

\author{Glenn R. Myers}
 \affiliation{Department of Applied Mathematics, Research School of Physics, The Australian National University, Canberra, ACT 2601, Australia}

\author{Daishi Adams}
 \affiliation{Department of Applied Mathematics, Research School of Physics, The Australian National University, Canberra, ACT 2601, Australia}

\author{Daniele Pelliccia}
\affiliation{Instruments \& Data Tools Pty Ltd, Victoria 3178, Australia}



\author{David M. Paganin}
\affiliation{School of Physics and Astronomy, Monash University, VIC 3800, Australia}

\date{\today}

\begin{abstract}

Classical ghost imaging is a computational imaging technique that employs patterned illumination. It is very similar in concept to the single-pixel camera in that an image may be reconstructed from a set of measurements even though all imaging quanta that pass through that sample are never recorded with a position resolving detector. The method was first conceived and applied for visible-wavelength photons and was subsequently translated to other probes such as x rays, atomic beams, electrons and neutrons. In the context of ghost imaging using penetrating probes that enable transmission measurement, we here consider several questions relating to the achievable signal-to-noise ratio (SNR). This is compared with the SNR for conventional imaging under scenarios of constant radiation dose and constant experiment time, considering both photon shot-noise and per-measurement electronic read-out noise. We show that inherent improved SNR capabilities of ghost imaging are limited to a subset of these scenarios and are actually due to increased dose (Fellgett advantage). An explanation is also presented for recent results published in the literature that are not consistent with these findings.

\end{abstract}

\maketitle


\section{Introduction}


Transmission imaging and diffraction with penetrating probes such as x rays \cite{paganin2006}, neutrons \cite{NeutronImaghingAndItsApplications}, and electrons \cite{CowleyBook} enables one to see inside an object (through radiography \cite{RadiographyBook}, computed tomography \cite{KakSlaneyBook,Natterer}, crystallography \cite{HammondCrystallographyBook}, ptychography \cite{Rodenburg2008}, etc.), in a non-destructive manner and potentially {\it in-situ} or {\it operando} \cite{Pietsch2016,Rack2020}. Important understanding of specimens may be obtained through these imaging modalities, that cannot be accessed through any other technique, at scales ranging from angstroms up to kilometers and beyond. Often, particularly in a biological-imaging setting, the radiation dose imparted to these specimens is an important consideration \cite{BushbergBook}. For living specimens, such as in medical and pre-clinical x-ray imaging, the probe may be carcinogenic \cite{Linet2012} and a trade-off must be made between the information gained and the probability of cancer developing. For excised samples, excessive dose can cause structural damage distorting the resulting images and thereby limiting their usefulness. How dose can be minimized is an important question to ask in this context. The same question is important in low-fluence settings \cite{Gureyev2016} where a paucity of imaging quanta necessitates low dose, not necessarily on account of sample damage, but rather on account of faint sources and/or short image-acquisition times; ultrafast imaging \cite{Olbinado2017} and imaging using faint sources \cite{Camara2008,Constable2010} falls into this category of imaging problems for which dose-reduction is also an important consideration.  

Ghost imaging (GI) is an unconventional imaging technique that has the potential to address this question. Ghost imaging was first developed in the field of quantum optics utilizing entangled photons \cite{Pittman1995optical}. In one arm of the ghost-imaging experiment the interaction of photons with an object of interest is recorded, while in the other arm, the position of one photon in entangled-photon pairs is  recorded. Neither set of measurements alone can yield an image of the object, and it is only through their correlation that an image emerges. Such ``ghost images'' can be produced with extremely low numbers of photons \cite{Morris2015, Moreau2017}. It was subsequently realized that only the correlation property is required of the photon pairs and, based upon this observation, classical ghost imaging was designed through the use of patterned illumination \cite{bennink2002two, shapiro2008computational}. Classical ghost imaging has now been translated to penetrating probes such as x rays \cite{pelliccia2016experimental, Yu2016fourier}, electrons \cite{li2018electron}, and neutrons \cite{kingston2020neutron}. The initial work with x rays in 1D has now been extended to 2D radiography by \citet{zhang2018table} and \citet{pelliccia2018towards}, as well as 3D computed tomography by Kingston {\em{et al.}}~\cite{KingstonOptica2018,KingstonIEEE2019}. In light of all of the above recent advances, it is timely to explore the possibilities of dose-reduction through ghost imaging.

The dose-reduction capabilities of ghost imaging seem to be poorly understood in the literature, with some inconsistency between theoretical predictions and experimental results. Gureyev {\it et al}.~\cite{gureyev2018} showed that for the signal-to-noise ratio from GI to match that for direct imaging with a pixelated detector, the number of measurements must be very low. To achieve a high SNR with only a few measurements would require either: (1) that the object image be strongly correlated with only a small number of patterned illuminations (with the specific patterns known {\it a priori}), or (2) that the object image be extremely sparse in some representational basis such that {\it compressed sensing} (CS) can be exploited. Lane and Ratner \cite{lane2020advantages} showed that the Fellgett (or multiplex) advantage applies to ghost imaging and provides superior SNR for a given experiment time when noise is dominated by a per-measurement read-out noise independent of signal. However, this SNR improvement is due to increased dose rather than improved utilization of dose. Lane and Ratner also showed that GI (and multiplexing in general) provides no advantage under photon shot-noise. They concluded that to reduce dose, GI must be combined with CS or related concepts. Ceddia and Paganin reached similar conclusions \cite{Ceddia2018}. Recent experimental results in the literature \cite{zhang2018table} have demonstrated x-ray GI with ultra-low dose that is stated to be superior to direct imaging; this seems to suggest that reduced-dose is an inherent property of GI. However, we believe these results can be explained by the absence of a shutter during detector read-out under continuous wave x-ray illumination (as will be detailed herein). 

In this paper we wish to re-iterate and expand upon the theoretical predictions of SNR from classical GI in settings where one is given no previous knowledge regarding the object being imaged. A key point, here, is that in the absence of any {\it a priori} knowledge regarding the object, reduced-dose is not an inherent property of classical GI. The power of ghost imaging (and computational imaging in general) for dose-reduction lies in the ability to capitalize on {\it a priori} knowledge. We do not need to measure object properties that are given, and seek to only make sufficient measurements to identify the differences from an expected result. Our focus here is therefore on classical ghost imaging since here this ability is encapsulated in either the basis of measurement (i.e., the illumination patterns employed) or the representational basis in image reconstruction in which the image is assumed sparse. Quantum ghost imaging (QGI), (first realised with X-rays by Schori et al. \cite{schori2018ghost}), while almost optimal in imaging efficiency, is more similar to direct imaging in this context since it does not utilize illumination masks and no image reconstruction is required. Owing to its inherent imaging efficiency, QGI can almost certainly provide a dose reduction, however, in our view it is likely to be very limited; this point, while important, lies beyond the scope of the present paper and therefore will not be further explored here.

The advantages and disadvantages of employing bucket detectors in ghost imaging over pixelated detectors are an important consideration in this paper. Since spatial resolution is not a consideration, bucket detectors typically have a higher detective quantum efficiency (DQE). Since there is no concept of spatial resolution, increasing signal spread has no effect. One can then increase the interaction lengths (for example by using thicker scintillating materials) to increase the degree of X-ray interaction (although, in the case of scintillators, the conversion of X-ray energy to signal may be reduced). Conversely efficiency becomes an important trade-off for smaller and smaller pixel sizes in a position sensitive detector. The electronic read-out noise may also be lower for bucket detectors compared with pixelated detectors. Other considerations include pixel cross-talk, (or charge sharing), and the effects of X-ray scatter on measurements. Both these factors affect pixelated detectors but do not apply to bucket detectors. For analysis and simulations throughout this paper we will assume the worst case scenario for GI by ignoring these effects and assuming equivalent DQE and electronic noise for both bucket detectors and pixelated detectors.

The remainder of the paper proceeds as follows. Section~\ref{sec:background} outlines classical ghost imaging and its properties. The point-spread-function of GI is derived and used to determine the required normalization scale for the conventional ghost imaging reconstruction formula. In Sec.~\ref{sec:SNR} the SNR of GI as a function of experiment parameters is studied. Following this, GI is compared with conventional imaging in Sec.~\ref{sec:giComparisonWithConvImaging}; the Fellgett advantage under per-measurement Gaussian noise is demonstrated, and it is then shown that GI provides no advantage under Poisson noise. A potential explanation for the contradictory GI-dose-reduction experimental results in Zhang~{\it et~al}.~\cite{zhang2018table} is presented in Sec.~\ref{sec:zhangExpl}, along with validation by simulation. This is followed by a discussion drawing from the results together with an indication of some possible directions for future research, in Sec.~\ref{sec:disc}. Concluding remarks are given in Sec.~\ref{sec:conc}.

\section{Classical ghost imaging}
\label{sec:background}

\begin{figure}[ht!]
\centering
\includegraphics[width=\linewidth]{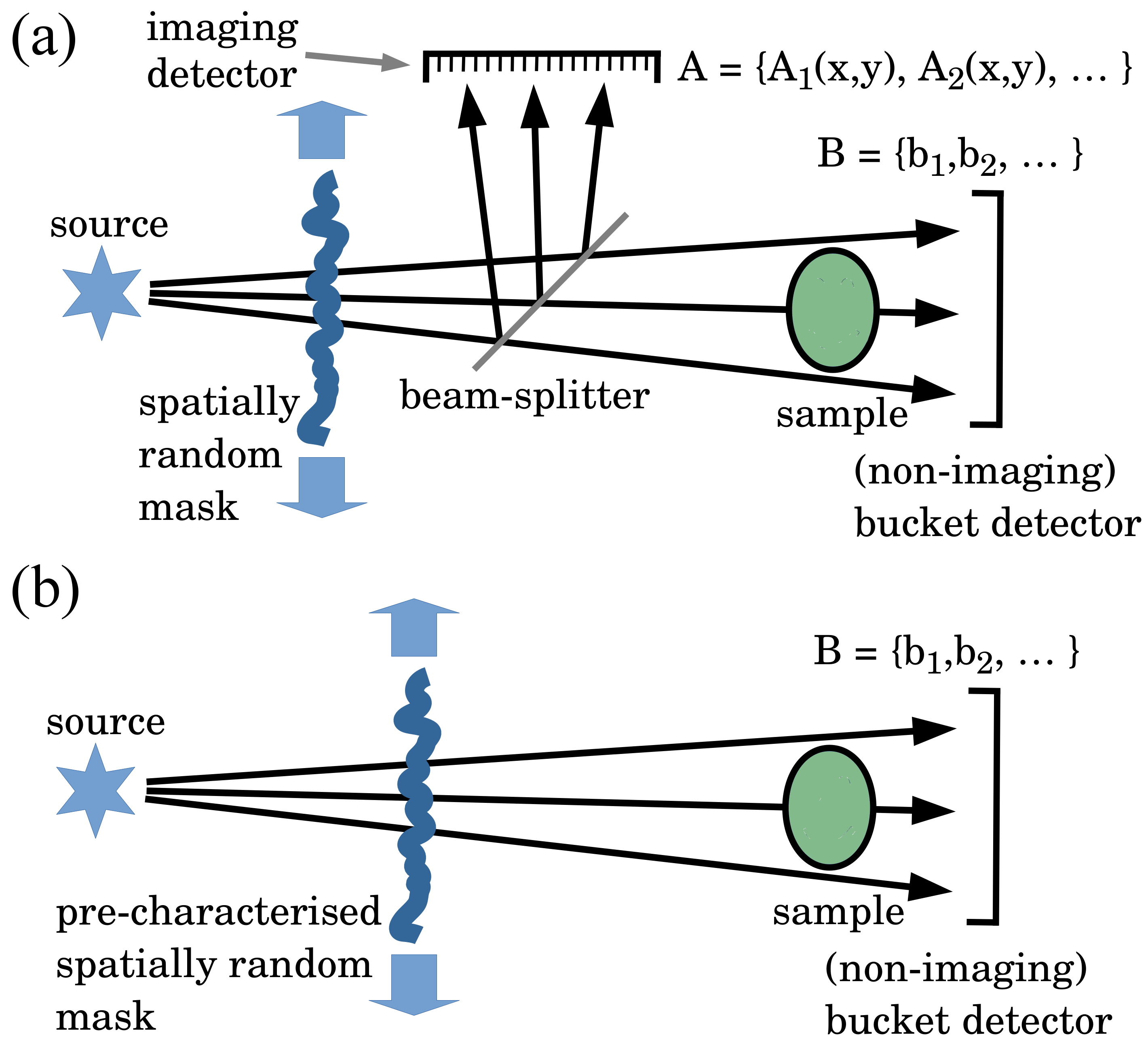}
\caption{Variations of the experimental setup to achieve classical ghost imaging. (a) Traditional setup using a beam-splitter for simultaneous measurement of illumination patterns, $A$, and bucket signals, $B$. (b) Computational set up where the illumination patterns, $A$, are first recorded, then the bucket signals, $B$, are measured with the repeated set of mask positions. \label{fig:expSetup}}
\end{figure}

While both quantum-mechanical and classical variants of the ghost imaging methodology exist, here and henceforth we restrict consideration to the classical case only, using a set of patterned illuminations. A key feature of GI is that all imaging quanta, that pass through the object being imaged, are never registered using a position-sensitive detector. Instead, they are recorded using a single-pixel detector, referred to as a ``bucket,'' to yield the total transmission of the sample \cite{katz2009compressive, Bromberg2009ghost} for a given patterned illumination. Imaging quanta that never pass through the object are measured using a position-sensitive detector, to give the set of illumination patterns (or speckle maps \footnote{Throughout this paper, we use the term ``speckle'' to simply refer to patterned intensity distributions. These distributions may be spatially-random or deterministic. This usage of ``speckle'' differs markedly from the common usage which equates ``speckle'' with ``fully developed coherent speckle''.}) that have no relation to the object being imaged. Note that the spatial distribution of these illumination patterns may arise due e.g.~to intrinsic beam characteristics such as the partially coherent character of an optical beam, or alternatively, patterned illumination may be imposed e.g.~by transversely scanning a spatially random or deterministic mask. A schematic depiction of typical experimental GI set-ups is presented in Fig.~\ref{fig:expSetup}. The speckle maps may be recorded simultaneously with the bucket measurements using a beam-splitter as shown in Fig.~\ref{fig:expSetup}(a). Controlled and repeatable speckle maps may be pre-recorded (or even known {\it a priori}) as shown in Fig.~\ref{fig:expSetup}(b).  The latter variant is referred to as {\it computational} ghost imaging \cite{shapiro2008computational}. When the  process described above is repeated many times, each with a different spatial distribution of the illumination, one obtains an ensemble of position-resolved illumination patterns, $A$, that may be statistically independent or deterministically orthogonal, together with a set of corresponding scalar bucket signals, $B$.

Given a set of $J$ patterned illuminations, $A = \{A_1(x,y), A_2(x,y), \ldots, A_J(x,y)\}$, where $x$ and $y$ are detector-plane coordinates \footnote{Here $x$ and $y$ are discrete variables that map to the pixel positions of the speckle maps recorded by the position-sensitive detector.}, the set of bucket signals measured, $B = \{b_1, b_2, ..., b_J\}$, can be modeled as the cross-correlations of $A$ with the transmission image $T(x,y)$ of the object of interest, i.e.,
\begin{equation}\label{eq:xc}
    b_j = \sum_x \sum_y A_j(x,y) T(x,y).
\end{equation}

While neither $A$ nor $B$ considered in isolation will enable a transmission image of the object to be formed, the {\it correlation} between these datasets enables a ghost image to be reconstructed \cite{katz2009compressive, Bromberg2009ghost}. Classical ghost imaging is then by necessity a {\it computational imaging} technique since no image can be formed directly from the measurements. The traditional ghost image recovery equation presented, generally without proof, in the literature (e.g., \cite{katz2009compressive,Bromberg2009ghost}) is the following:
\begin{equation}\label{eq:adjoint}
\widehat{T}(x,y) = \sum_j A_j(x,y) ( b_j - \langle b \rangle).
\end{equation}
Throughout the paper we will use $\langle . \rangle$ to denote the mean value in dimension $j$, i.e., $\langle \eta \rangle = \frac{1}{J} \sum_j \eta_j$; thus here $\langle b \rangle$ denotes the mean bucket signal. It is shown in Appendix \ref{app:a} that this can be derived as the adjoint of a mean-corrected form of the cross-correlation equation (Eq.~(\ref{eq:xc})). While in many cases the recovered image, $\widehat{T}(x,y)$, may resemble the original object transmission, $T(x,y)$, they may not be equal for several reasons:
\begin{enumerate}
    \item A scale factor, $\gamma$, is introduced such that $\langle \widehat{T} \rangle = \gamma \langle T \rangle$. While it is possible that $\gamma=1$, for example if the illumination patterns form an orthonormal basis, in general this scale factor will not be equal to unity. 
    \item The adjoint operation (scaled by $\gamma^{-1}$) is only equal to the inverse for unitary operators.  In our context, the adjoint will equal the inverse if the illumination patterns form an orthogonal basis (e.g., the set of 2D binary patterns based on the Hadamard transform). It will also be shown later that for a non-orthogonal basis, the scaled adjoint approaches the inverse as $J \rightarrow \infty$.
    \item If the set of illumination patterns has a constant total transmission, i.e., $\sum_x \sum_y A_j(x,y) = k ~ \forall ~ j \in [1,J]$, the mean object transmission is lost so that $\langle \widehat{T} \rangle = 0$. This condition often arises when an orthogonal scanning mask is employed (such as a pin-hole or a uniformly redundant array). 
\end{enumerate}

Each of these issues will be addressed in subsequent subsections, however, prior to this it is beneficial to understand the Green's functions associated with ghost imaging \cite{gureyev2018}, as well as a shift invariant point spread function (PSF) representing the expected Green's functions.

\subsection{Point spread function}
\label{sec:psf}

The ghost imaging process is modeled by combining Eq.~(\ref{eq:xc}) with Eq.~(\ref{eq:adjoint}) and rearranging to obtain \footnote{To achieve this result we have used the fact that $\sum_j \langle A(x,y) \rangle \tilde{A}_j(x,y) = 0$.}:
\begin{equation}\label{eq:xcAdj}
\widehat{T}(x,y) = \sum_{x'}\sum_{y'} T(x',y') \sum_j \tilde{A}_j(x',y')\tilde{A}_j(x,y),
\end{equation}
where $\tilde{A}_j(x,y) = A_j(x,y) - \langle A(x,y)\rangle$. The corresponding Green's function $G_{(x^*,y^*)}$ describes the point spread effects of the ghost imaging process about the point $(x^*,y^*)$. A reconstructed ghost image can be decomposed into a sum of the set of Green's functions. $G_{(x^*,y^*)}$ is determined by inserting a Dirac delta function at $(x^*,y^*)$, i.e.~setting $T(x',y') = \delta(x'-x^*, y'-y^*)$, and applying Eq.~(\ref{eq:xcAdj}) as follows:
\begin{eqnarray}
    G_{(x^*,y^*)} (x,y)
    & = & \sum_{x'} \sum_{y'} \delta(x'-x^*, y'-y^*) \nonumber\\
    && \quad \sum_j \tilde{A}_j(x',y') \tilde{A}_j(x,y) \nonumber \\
    & = & \sum_j \tilde{A}_j(x^*,y^*) \tilde{A}_j(x,y).
\end{eqnarray}

\begin{figure}[ht!]
    \centering
    \begin{minipage}{0.4\linewidth}
    \centering
    \scriptsize{(a)}\\
    \includegraphics[width=0.95\linewidth]{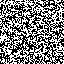}
    \end{minipage}%
    \begin{minipage}{0.4\linewidth}
    \centering
    \scriptsize{(b)}\\
    \includegraphics[width=0.95\linewidth]{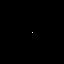}
    \end{minipage}\\
    \begin{minipage}{0.4\linewidth}
    \centering
    \scriptsize{(c)}\\
    \includegraphics[width=0.95\linewidth]{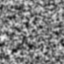}
    \end{minipage}%
    \begin{minipage}{0.4\linewidth}
    \centering
    \scriptsize{(d)}\\
    \includegraphics[width=0.95\linewidth]{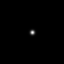}
    \end{minipage}
    \caption{Ghost imaging PSF determined using Eq.~(\ref{eq:GhostImagingPSF}) from a set of speckled illuminations. (a) An example $64 \times 64$ pixel random binary speckled illumination pattern. (b) The ghost imaging PSF obtained using 65,536 random patterns. (c) An example blurred speckled illumination generated by convolving the image in panel (a) with a Gaussian having a standard deviation $\sigma = 1.0$px. (d) The ghost imaging PSF obtained using 65,536 blurred random patterns.}
    \label{fig:PSF}
\end{figure}

The expected Green's function is described by a shift invariant PSF found as the average over all registered (or co-aligned) Green's functions. Registration is achieved by shifting each $G_{(x^*,y^*)}$ to be about a common point, e.g., about $(x,y) = (0,0)$ as  $G_{(x^*,y^*)}(x+x^*,y+y^*)$. The PSF is then found as follows:
\begin{eqnarray}\label{eq:GhostImagingPSF}
    \mbox{PSF}(x,y)
    & = & \frac{1}{n^2}\sum_{x^*}\sum_{y^*} G_{(x^*,y^*)} (x+x^*,y+y^*)\nonumber \\
    & = & \frac{1}{n^2}\sum_{x^*}\sum_{y^*} \sum_j \tilde{A}_j(x^*,y^*) \tilde{A}_j(x+x^*,y+y^*) \nonumber \\
    & = & \frac{J}{n^2} \langle \tilde{A} \circ \tilde{A} \rangle (x,y),
\end{eqnarray}
where $\circ$ denotes 2D correlation. The PSF (or expected Green's function) is therefore  directly proportional to the mean auto-correlation of speckle patterns. As a simple simulated example to exemplify this process, the PSFs for two sets of $J = 65,536 \gg n^2$ speckle patterns generated as $64 \times 64$ pixel images are presented in Fig.~\ref{fig:PSF}.

\begin{figure}[ht!]
    \centering
    \begin{minipage}{0.4\linewidth}
    \centering
    \scriptsize{(a)}\\
    \includegraphics[width=0.95\linewidth]{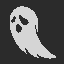}
    \end{minipage}%
    \begin{minipage}{0.4\linewidth}
    \centering
    \scriptsize{(b)}\\
    \includegraphics[width=0.95\linewidth]{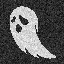}
    \end{minipage}\\
    \begin{minipage}{0.4\linewidth}
    \centering
    \scriptsize{(c)}\\
    \includegraphics[width=0.95\linewidth]{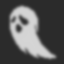}
    \end{minipage}%
    \begin{minipage}{0.4\linewidth}
    \centering
    \scriptsize{(d)}\\
    \includegraphics[width=0.95\linewidth]{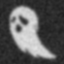}
    \end{minipage}
    \caption{Demonstration of ghost imaging as the convolution of the original transmission image with a point-spread-function (PSF) determined from the set of speckled illuminations. (a) The original $64 \times 64$ pixel image. (b) Result from ghost imaging using 65,536 random binary speckled illumination patterns (see example speckle pattern in Fig.~\ref{fig:PSF}(a)). (c) The original image in panel (a) blurred by convolution with the ghost imaging PSF in Fig.~\ref{fig:PSF}(d). (d) Result from ghost imaging using the 65,536 random binary speckled illumination patterns used in panel (b), blurred by convolution with a Gaussian with $\sigma = 1$px (see example speckle pattern in Fig.~\ref{fig:PSF}(c)).}
    \label{fig:giPsfDemo}
\end{figure}

Assuming spatial statistical stationarity (see e.g., \cite{Goodman2015}), we approximate each $G_{(x',y')}(x,y)$ as $\mbox{PSF}(x-x',y-y')$ and Eq.~(\ref{eq:xcAdj}) becomes:
\begin{eqnarray}\label{eq:giAsConv}
    \widehat{T}(x,y)
    & = & \sum_{x'}\sum_{y'} T(x',y') G_{(x',y')}(x,y)\nonumber \\
    & \approx & \sum_{x'}\sum_{y'} T(x',y') \mbox{PSF}(x-x',y-y')\nonumber \\
    & = & \{ T \ast \mbox{PSF} \} (x,y),
\end{eqnarray}
where $\ast$ denotes 2D convolution. This approximation has been demonstrated in Fig.~\ref{fig:giPsfDemo} for the two speckle patterns presented in Fig.~\ref{fig:PSF}. The results of $\{ T \ast \mbox{PSF} \}$ are presented in Fig.~\ref{fig:giPsfDemo}(a) and (c) with the corresponding ghost images given in Fig.~\ref{fig:giPsfDemo}(b) and (d). The resolution of the ghost images is limited to the width of the PSF (as predicted by Eq.~(\ref{eq:giAsConv})) \cite{pelliccia2018towards, gureyev2018}; we also observe some additional background artefacts that appear as noise. These artefacts disappear as $J \rightarrow \infty$.

\subsection{Normalization scale for the adjoint operator}
\label{sec:adjScale}

It was shown in Eq.~(\ref{eq:giAsConv}) that the ghost imaging process can be approximated as the convolution of the object transmission image with the ghost imaging PSF. From this observation we see that the multiplicative scale introduced by the ghost imaging process such that $\widehat{T}(x,y) = \gamma T(x,y)$ is found as the integral of the PSF with respect to transverse position, i.e.,
\begin{equation}\label{eq:gamma}
    \gamma = \sum_x \sum_y \mbox{PSF}(x,y) = \frac{J}{n^2} \sum_x \sum_y \left< \tilde{A} \circ \tilde{A} \right> (x,y).
\end{equation}
Here, the scale required to normalize the adjoint operation in Eq.~(\ref{eq:adjoint}) is $\gamma^{-1}$. By way of demonstration, Appendix \ref{app:b} presents the calculation of $\gamma$ for the examples presented in Fig.~\ref{fig:PSF} and Fig.~\ref{fig:giPsfDemo}. For independent random patterned illuminations, $\gamma = J \sigma_A^2$ where $\sigma_A^2$ is the variance of the speckle patterns, while for the random patterns blurred by a 2D Gaussian function (with standard deviation $\sigma_g$), $\gamma = 4 \pi J \sigma_g^2 \sigma_A^2$.

\subsection{Landweber iteration for a general set of illumination patterns}
\label{sec:landweber}

Given an orthogonal set of illumination patterns the adjoint operator scaled by $\gamma^{-1}$ is the inverse operation, although this is not the case for a general set of illumination patterns \footnote{An indication, of the flexibility in choice of illumination patterns, is given by a recent paper on computational GI using illumination patterns given by successive greyscale frames of a Charlie Chaplin movie \cite{GI-with-Charlie-Chaplin-movie2020}.  The key point, here, is that the ensemble of illuminations be linearly independent of one another \cite{gureyev2018}.}. Let us present ghost imaging in matrix notation where $\mathbf{t}$ is the image $T(x,y)$ represented as a vector, $\tilde{\mathbf{A}}$ is a matrix with each row formed as the transpose of $\tilde{A}_j(x,y)$ as a vector, and $\tilde{\mathbf{b}}$ is the vector of mean-corrected bucket measurements, $\tilde{b}_j = b_j - \langle b \rangle$. The mean-corrected form of Eq.~(\ref{eq:xc}) then becomes $\tilde{\mathbf{b}} = \tilde{\mathbf{A}}\mathbf{t}$ and the normalized adjoint in Eq.~(\ref{eq:adjoint}) becomes
\begin{equation}\label{eq:adjMatrix}
    \widehat{\mathbf{t}} = \frac{1}{\gamma}\tilde{\mathbf{A}}^\mathrm{T}\tilde{\mathbf{b}}.
\end{equation}
For orthogonal mean-corrected speckle patterns, $\tilde{\mathbf{A}}^\mathrm{T} = \gamma \tilde{\mathbf{A}}^{-1}$ so Eq.~(\ref{eq:adjMatrix}) is the inverse. However, this is not true in general and the Moore-Penrose inverse $\left( \tilde{\mathbf{A}}^\mathrm{T}\tilde{\mathbf{A}} \right)^{-1}$ is employed as follows:
\begin{equation}
    \widehat{\mathbf{t}} = \left( \tilde{\mathbf{A}}^\mathrm{T}\tilde{\mathbf{A}} \right)^{-1} \tilde{\mathbf{A}}^\mathrm{T}\tilde{\mathbf{b}}.
\end{equation}
For large images, computing the Moore-Penrose inverse may be prohibitive and the following Landweber algorithm that iteratively updates the current estimate, $\widehat{\mathbf{t}}^i$, to $\widehat{\mathbf{t}}^{i+1}$ may be preferable:
\begin{equation}\label{eq:landweber}
    \widehat{\mathbf{t}}^{i+1} = \widehat{\mathbf{t}}^i + \frac{\alpha}{2\gamma}\left( \tilde{\mathbf{A}}^\mathrm{T} (\tilde{\mathbf{b}} - \tilde{\mathbf{A}}\widehat{\mathbf{t}}^i)  \right),
\end{equation}
where $\alpha \in (0,1]$ is a regularization term. Note that this algorithm is maximum likelihood assuming uniform Gaussian noise. A modified form may be preferable if noise is dominated by photon shot-noise modeled as a Poisson distribution. 

\subsection{Correction for illumination patterns with constant total transmission}
\label{sec:constMeanMasks}

While 2D Hadamard masks can be fabricated as stencils and do provide an orthogonal set of illumination patterns, they are extremely inefficient since each mask is only used once. To generate an $n \times n$ pixel ghost image using Hadamard masks requires fabricating $n^2$ binary stencils with $n \times n$ elements. A far more efficient mask is one that is orthogonal under translation. Such masks include pin-hole masks, uniformly redundant arrays (URA, e.g., \cite{Gottesman1989newFamily}) and masks based on the finite Radon transform (FRT, e.g., \cite{cavy2015construction}). To generate an $n \times n$ pixel ghost image in this case requires a single stencil with $(2n-1) \times (2n-1)$ elements (fabricated such that $M(x,y) = M(x+n,y+n)$). The ghost imaging experiment illuminates only $n \times n$ elements of this stencil per measurement and is translated to the $n^2$ available positions to record the set of bucket signals.

These sets of illumination patterns have a constant total transmission, i.e., $\sum_x \sum_y A_j(x,y) = k ~ \forall ~ j \in [1,J]$. In this case the mean object transmission is lost in ghost image reconstruction, so that $\langle \widehat{T} \rangle = 0$. In these cases the mean, $\langle T \rangle = \langle b \rangle / k$, can be added to $\gamma^{-1}\widehat{T}$.

\section{Signal-to-noise ratio study}
\label{sec:SNR}

Having addressed the deficiencies of ghost image reconstruction we are now in a position to investigate the signal-to-noise ratio, where the {\it noise} is defined as the root-mean-square error (RMSE), i.e.,
\begin{equation}\label{eq:rmse}
    \mbox{RMSE}(\widehat{T},T) = \sqrt{ \frac{1}{n^2}\sum_x \sum_y \left(\widehat{T}(x,y) - T(x,y)\right)^2 }.
\end{equation}
Here we first establish the {\it signal-to-noise ratio} of ghost imaging as a function of the imaging parameters. We then compare the SNR of ghost imaging with that of (a) a scanning probe and (b) direct imaging with a position sensitive detector. This underpins the overall objective of our paper, which is to identify the capabilities of GI to achieve the same SNR or contrast-to-noise ratio (CNR) as these more conventional imaging techniques while subjecting the object under investigation to minimal radiation dose.

Since we are working with simulations, {\it noise} can be specified as the error in image reconstruction, as quantified by RMSE$(\widehat{T},T)$. The definitions of {\it signal} or {\it contrast} are functions of object transmission, (i.e., proportional to the mean, $\mu_T$, and standard deviation, $\sigma_T$, respectively), and independent of imaging parameters. It is therefore not important in this study on the effect of imaging parameters. In a manner similar to peak signal-to-noise ratio (PSNR) often used for evaluation of lossy codec performance  \footnote{Note that one can consider ghost imaging as a codec (coder-decoder): The ghost imaging measurements encode the original image of the object as coefficients in a space where the illumination patterns form the basis vectors, and the ghost image reconstruction process decodes this data back to image space.}, we will use the maximum possible signal (the case of 100\% transmission) as the numerator and investigate SNR defined simply as $\mbox{SNR}(\widehat{T},T) = 1 / \mbox{RMSE}(\widehat{T},T)$.

In what follows, we consider monochromatic, parallel-beam illumination with photon flux $\mathcal{B}$ photons/s/mm$^2$. Speckle patterns are generated by illuminating masks fabricated with $n \times n$ elements of area $\Delta^2$mm$^2$, with transmission of element $(x,y)$ of mask $j$ described by $A_j(x,y)$. With mask $j$ in place, the number of photons incident at pixel $(x,y)$ of the object after an exposure of duration $t_0$ is described by $\mathcal{B} A_j(x,y) t_0 \Delta^2$. Given $J$ patterned illuminations, the total exposure time is $J t_0$ and we define the total number of incident photons per pixel from a ghost imaging experiment as $\mathcal{P} = \mathcal{B} J t_0 \Delta^2$. Note that throughout this paper we have assumed a uniform incident illumination (or flatfield). In what follows, we consider ensembles of speckle patterns that produce a PSF that is a delta function with integral $J\sigma_A^2$.

In the high-photon-flux (or noise-free) limit, given a set of $J$ independent, random masks with transmission, $A_j(x,y)$, the RMSE (as derived in Appendix \ref{app:c}) is as follows:
\begin{equation}\label{eq:snrNoiseFreeRandom}
    \mbox{SNR}_0(T,\widehat{T}) = \frac{1}{\mbox{RMSE}_0(T,\widehat{T})} \approx \sqrt{ \frac{ J }{ n^2 (\mu^2_T + \sigma^2_T) }  }.
\end{equation}
This quantifies the natural dependencies that (1) for a given original image, the SNR is proportional to the square root of the number of masks employed, $J$ --- a dependence that arises from the random-basis character \cite{Gorban2016, Ceddia2018} of the ensemble of illuminating speckle maps; (2) For fixed $J$, the SNR becomes lower as the original signal becomes larger (either mean or standard deviation increase). This is a generalization of the finding in Ref.~\cite{paganin2019writing} for binary images that SNR is inversely proportional to the square-root of the number of non-zero elements; for binary images $\mu^2_T + \sigma^2_T = \mu_T$. We note that this analysis of the adjoint operation provides an estimate of the lower limit on SNR. Reconstruction through the Moore-Penrose inverse or Landweber iteration (via Eq.~(\ref{eq:landweber})) can potentially improve SNR up to the equivalent to that for a set of orthogonal masks.

Given a scanning mask that is orthogonal under translation (such as a uniformly redundant array \cite{Gottesman1989newFamily} or that generated through the finite Radon transform \cite{cavy2015construction}), the RMSE (as derived in Appendix \ref{app:c}) is as follows:
\begin{equation}\label{eq:snrNoiseFreeOrtho}
    \mbox{SNR}^\perp_0(T,\widehat{T})
    = \frac{1}{\textrm{RMSE}^\perp_0(\widehat{T},T)}
    \approx \sqrt{\frac{ n^2 }{ (n^2 - J) \sigma^2_T}},
\end{equation}
for $1 \le J < n^2$. While this equation becomes a function of the number of masks missing, $n^2-J$, observe that SNR increases as the number of masks missing decreases; both this, Eq.~(\ref{eq:snrNoiseFreeOrtho}) and SNR for the random-mask case, Eq.~(\ref{eq:snrNoiseFreeRandom}), are monotonically increasing functions of the number of masks, $J$. (Simulations demonstrating the validity of these noise-free SNR estimations have been plotted in Fig.~\ref{fig:snrStudyNoiseFreePlots} in Appendix \ref{app:c}.)

The above SNR observations for the ``noise free'' case could be considered as the signal-to-artefact ratio (SAR). Let us now investigate the effect of incorporating experiment noise into the measured bucket signals. We will consider two  extreme cases: (1) photon shot-noise modeled as a Poisson distribution scaled by $\sigma_p$, and (2) per-measurement detector noise modeled as a Gaussian distribution with uniform standard deviation, $\sigma_m$. Experimental noise can typically be modeled as a weighted combination of Poisson and Gaussian distributions (see \cite{nuyts2013modelling}). The RMSE for both types of noise is derived in Appendix \ref{app:c} in the limit where there are no artefacts from GI, (i.e., $J \gg n^2$ for the random mask case or $J=n^2$ for orthogonal masks). In this limit, SAR $\rightarrow \infty$ and SNR approaches more ``conventional'' imaging definitions. For photon-shot noise:
\begin{equation}\label{eq:snrPoisson}
\textrm{RMSE}_p(\widehat{T},T) = \sqrt{ \frac{ \sigma_p^2 \mu_A \mu_T n^2 }{ \mathcal{P} \sigma_A^2 } },
\end{equation}
where $\mu_A$ is the mean transmission of the speckle masks. In the same limit with per-measurement electronic read-out noise:
\begin{equation}\label{eq:snrGauss}
\mathrm{RMSE}_m(\widehat{T},T) = \sqrt{ \frac{ J \sigma_m^2 }{ \mathcal{P}^2 \sigma_A^2 } }.
\end{equation}
Equation~(\ref{eq:snrNoiseFreeRandom}) for a set of random masks in the presence of noise then becomes:
\begin{widetext}
\begin{eqnarray}\label{eq:snrRandom}
    \textrm{SNR}(\widehat{T},T)
    & = & 1 / \sqrt{ RMSE^2_0(\widehat{T},T) + RMSE^2_p(\widehat{T},T) + RMSE^2_m(\widehat{T},T}) \nonumber \\
    & = & 1 / \sqrt{ \frac{ n^2 (\mu^2_T + \sigma^2_T)}{J} + \frac{ \sigma_p^2 \mu_A \mu_T n^2 }{\mathcal{P} \sigma_A^2 } + \frac{ J \sigma_m^2}{\mathcal{P}^2 \sigma_A^2 } }.
\end{eqnarray}
\end{widetext}
The expansion of $\mbox{SNR}^\perp_0$ for orthogonal masks (Eq.~(\ref{eq:snrNoiseFreeOrtho})) proceeds similarly.

A suite of simulations is presented in Appendix \ref{app:d}, to both: (1) validate the above equations, and (2) demonstrate the effect of each parameter on the recovered ghost image SNR given a constant exposure time per illumination pattern of $t_0$ seconds. As outlined in Appendix \ref{app:d}, $n \times n = 64 \times 64$ pixel uniformly random arrays, with $T(x,y) \in [0,1)$, were used as 1mm$^2$ transmission images of the object. The pixel-pitch is therefore $1/n$mm and the pixel area is $1/n^2$mm$^2$. A parallel x-ray beam is assumed with a set of 1mm$^2$ speckle masks with a resolution of $2/n$mm creating the patterned illuminations for ghost imaging. Given an incident photon flux of $\mathcal{B} = 4.1 \times 10^5$ photons/s/mm$^2$ and an exposure time of $t_0 = 0.01$s per bucket measurement, the total experiment time is then $\tau = J t_0$ seconds and $\mathcal{P} = \mathcal{B} J t_0 / n^2$. The resulting speckled illuminations are represented by $n \times n$ pixel random binary images with $\mathcal{B} t_0 A_j(x,y) / n^2$ photons per pixel for $j \in [1,J]$. Poisson and Gaussian distributions were used to simulate bucket measurements with photon shot-noise and per-measurement electronic noise respectively. For Poisson noise $\sigma_p = 1$, and for the case of Gaussian noise, $\sigma_m$ is set to yield similar noise levels to the Poisson case. Based on these plots and analyzing Eqs.~(\ref{eq:snrNoiseFreeRandom})--(\ref{eq:snrRandom}), we make the following observations:
\begin{itemize}
    
    \item Increasing the image dimension, $n$, reduces the SNR in all scenarios for random masks. Given an orthogonal scanning mask, SNR reduces as $n$ increases for both photon shot-noise and in the high-photon-flux limit (or noise-free case), but has no effect on the per-measurement noise component of SNR;
    
    \item Increasing the mean mask transmission, $\mu_A$, has no effect with per-measurement noise, however, this reduces SNR for photon shot-noise since the noise is proportional to $\mu_A$ while the GI signal is proportional to the mask standard deviation, $\sigma_A$, and not $\mu_A$;
    
    \item When the mask standard deviation, $\sigma_A$, is increased, SNR increases. As was derived in the previous section, the magnitude of the signal is proportional to $\sigma_A^2$, thus the greater the contrast of the random masks (for a given mean), the better the GI SNR when noise is incorporated. The GI artefacts are also proportional to $\sigma_A^2$, so this parameter has no effect in the noise-free case;

    \item Increasing the incident illumination per pixel, $\mathcal{P} = \mathcal{B} J t_0 / n^2$, increases SNR for both photon shot-noise and per-measurement noise as they asymptote to the noise-free case. The former is proportional to $t_0$ and $\mathcal{B}$ while the latter is proportional to $t_0^2$ and $\mathcal{B}^2$. There is a ``law of diminishing returns'' once $\mathcal{P}$ is so large that the second and third terms (RMSE$_p$ and RMSE$_m$ respectively) under the square root become negligible compared to the first term (RMSE$_0$ or RMSE$^\perp_0$ depending on the type of masks employed).
    
\end{itemize}

\begin{figure}[ht!]
    \centering
    \begin{minipage}{0.8\linewidth}
    \centering
    \scriptsize{(a)}\\
    \includegraphics[width=\linewidth]{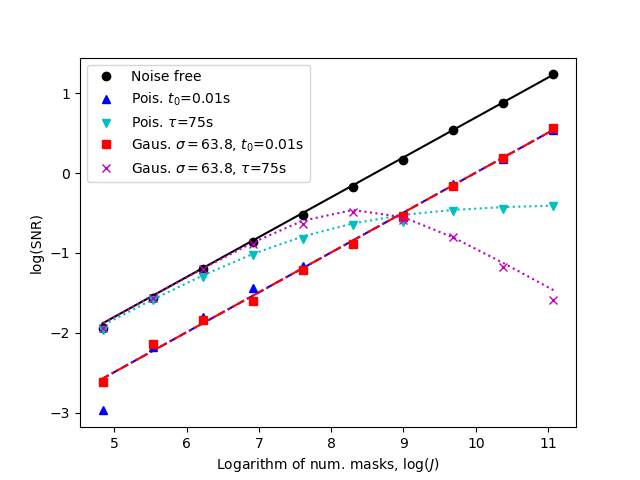}
    \end{minipage}\\
    \begin{minipage}{0.8\linewidth}
    \centering
    \scriptsize{(b)}\\
    \includegraphics[width=\linewidth]{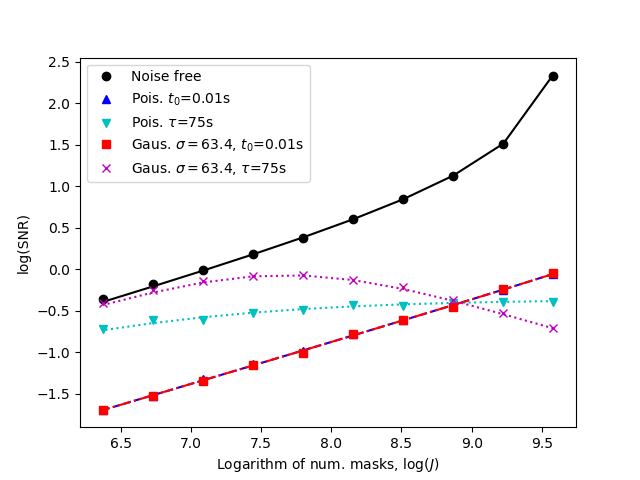}
    \end{minipage}
    \caption{Demonstration of the SNR of ghost imaging under Poisson and Gaussian noise models for (a) random masks, and (b) orthogonal masks, as a function of number of masks, $J$; simulations are run with either constant exposure time per mask, $t_0=0.01$s, or constant total experiment time, $\tau=82$s. Markers in the plot are simulation results while the lines connect the expected result according to Eq.~(\ref{eq:snrRandom}) and its orthogonal equivalent. Random masks and orthogonal finite Radon transform based masks were simulated and the ghost images have been recovered using the scaled adjoint given in Eq.~(\ref{eq:adjMatrix}).
    \label{fig:snrStudyConstTPlots}}
\end{figure}

Assuming a constant exposure time per measurement ($t_0 = 0.01$s in the simulations), the total dose on the object is increased with the number of masks, $J$. Consider the case where the total experiment time is constant ($\tau = 75$s in the following simulations); the available dose is then distributed among the $J$ measurements. SNR is affected in a different manner since the first observation above, assuming $\Xi \propto J$, no longer holds. The results of simulation demonstrating SNR as a function of $J$ have been presented in Fig.~\ref{fig:snrStudyConstTPlots}.  Here, three cases have been plotted: (1) the high-photon-flux limit (or noise-free case) where $t_0 = \tau = \infty$, (2) constant exposure time $t_0 = 0.01$s, and (3) constant total experiment time, $\tau = 82$s; for the latter two cases both photon shot-noise and per-measurement noise have been simulated.

We observe that, in this scenario, SNR will increase up to the point where artefacts (RMSE$_0$ or RMSE$^\perp_0$) are no longer the predominant contribution to total RMSE. For photon-shot noise we see that SNR seems to increase as $J$ is increased (with diminishing returns). This seems to be a parallel to the dose-fractionation theorem in computed tomography \cite{hegerl_zn_1976}. For per-measurement noise, no such fractionation exists; each extra measurement introduces additional error and a maximum SNR is reached for a particular value where (for the random masks) RMSE$_0$ matches RMSE$_m$:
\begin{equation}
    J_{\mbox{opt}} = \sqrt{ \frac{\mathcal{P}^2 n^2 \sigma^2_A (\mu_T^2 + \sigma_T^2) }{\sigma_M^2} } =  \frac{\mathcal{P} \, n \, \sigma_A \sqrt{\mu_T^2 + \sigma_T^2}
}{\sigma_M} .
\end{equation}
For the case in Fig.~\ref{fig:snrStudyConstTPlots}(a), we calculate $\ln (J_{\mbox{opt}}) = 7.9$. The analysis for orthogonal mask proceeds similarly, finding $J$ where RMSE$^\perp_0$ matches RMSE$_m$.

Now that the SNR for GI as a function of imaging parameters is understood, we can now compare performance with the more conventional imaging techniques of a scanning probe as well as direct imaging with a pixelated detector. This is explored in the following section.

\section{Comparison with conventional imaging}
\label{sec:giComparisonWithConvImaging}

Theoretical studies in the literature, such as those of Gureyev {\it et al}.~\cite{gureyev2018} and Lane and Ratner \cite{lane2020advantages}, have concluded that improved signal-to-noise ratio, or equivalently dose-reduction, is not a general benefit of ghost imaging. This is inconsistent with dose reduction of orders-of-magnitude that was reported in Zhang {\it et al}.~\cite{zhang2018table}; a potential explanation for this discrepancy is detailed in Sec.~\ref{sec:zhangExpl} below. However, there are instances where GI may be advantageous, e.g., under certain noise conditions, or given some {\it a priori} knowledge of the object. Later sections of this paper consider optimal mask design and image reconstruction for GI given {\it a priori} knowledge of the object. In this section we explore the performance of GI compared with that of the conventional imaging techniques: scanning probe imaging, and direct imaging with a pixelated detector under the noise conditions already explored (photon shot-noise and per-measurement electronic read-out noise).

Observe that conventional imaging can be considered a subset of ghost imaging: conceptually, a scanning probe is ghost imaging with a pin-hole mask, while direct imaging is a parallel scanning probe. The image produced by a {\it scanning probe} with area $\Delta^2$mm$^2$ translated in an $n \times n$ array of positions with step-size of $\Delta$mm and a dwell time of $t_0$s, is equivalent in dose and quality to a {\it direct image} using an $n \times n$ pixel detector with a pixel-pitch of $\Delta$mm and an exposure time of $t_0 / n^2$s. Therefore in what follows, we compare GI with a scanning probe given a total experiment time and compare GI with {\it direct imaging} given a total dose incident on the object.

\subsection{Detector electronic read-out noise}
\label{sec:gaussNoise}

Lane and Ratner \cite{lane2020advantages} observed that the Fellgett (or multiplex) advantage \cite{fellgett1949ultimate, fellgett1951theory} applies to GI under per-measurement noise. An improvement in SNR results when recording multiplexed measurements rather than direct measurements, when noise is dominated by detector noise. Multiplexing is traditionally performed in frequency (or energy), however, in GI multiplexing is in the spatial domain. The signal flux can be increased by measuring more regions of the object per measurement while noise remains constant.

\subsubsection{Comparison with scanning probe imaging (constant total experiment time, $\tau$)}
\label{sec:spVsGiGauss}

To compare GI performance, given $J$ measurements with a total experiment time of $\tau$s, with a scanning probe (SP) we assume that the same hardware is utilized for both experiments. An $n \times n$ pixel scanning probe image produced from a total exposure time of $\tau = Jt_0$s has a dwell time per pixel of $\tau / n^2$s. Each measurement has a standard deviation of $\sigma_m$, therefore we estimate SNR as $\mathrm{SNR}_{SP} = \mathcal{P} / n^2 \sigma_m$. Assuming that degradation from image reconstruction artefacts is negligible, the SNR from ghost imaging reduces to the reciprocal of Eq.~(\ref{eq:snrGauss}). Without knowledge of the object, to achieve negligible artefacts we therefore require $J \ge n^2$. We can then determine the relationship:
\begin{equation}\label{snrSpVsGiGauss}
\mathrm{SNR}_{SP} = \sqrt{ \frac{ J }{ n^4 \sigma_A^2 } } \textrm{SNR}_{GI}.
\end{equation}

\begin{figure}[ht!]
    \centering
    \includegraphics[width=\linewidth]{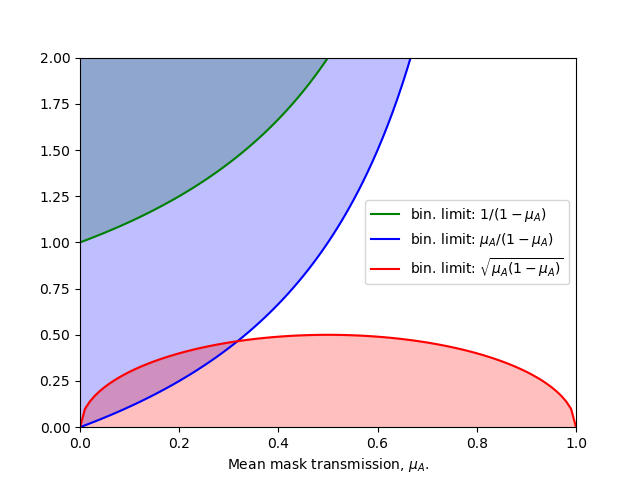}
    \caption{Several speckle mask properties as a function of the mean transmission, $\mu_A$. The area shaded red provides possible values that the standard deviation of mask transmission, $\sigma_A$, can assume with an upper bound defined by the binary mask case. The areas shaded green (resp.~blue) provide the possible values that $\mu_A/\sigma_A^2$ (resp.~$\mu_A^2/\sigma_A^2$) can assume with a lower bound defined by the binary mask case.}
    \label{fig:maskPropPlots}
\end{figure}

When considering the mask properties, we observe that the upper limit to $\sigma^2_A$ for a given $\mu_A$ is defined by the binary case, $A_j(x,y) \in \{0,1\}$, where $\sigma^2_A = \mu_A(1-\mu_A)$ (see Fig.~\ref{fig:maskPropPlots}); this gives $1/n^2 - 1/n^4 \le \sigma^2_A \le 0.25$. Therefore $\mathrm{SNR}_{SP} \le \textrm{SNR}_{GI}$, with equality being obtained only when using a pin-hole mask. So the scanning probe in this scenario can do no better than GI and typically $\mathrm{SNR}_{SP} \ll \textrm{SNR}_{GI}$. However, the total dose on the object for the scanning probe is just $\mathcal{P}$ while in ghost imaging the dose is $\mathcal{P} n^2 \mu_A$, i.e., $n^2 \mu_A$ times larger.

\begin{figure*}[ht!]
    \centering
    \begin{minipage}{0.08\linewidth}
    \centering
    \scriptsize{~}
    \end{minipage}%
    \begin{minipage}{0.15\linewidth}
    \centering
    \scriptsize{(a) Scanning probe imaging}
    \end{minipage}%
    \begin{minipage}{0.15\linewidth}
    \centering
    \scriptsize{(b) Ghost imaging, random masks, XC}
    \end{minipage}%
    \begin{minipage}{0.15\linewidth}
    \centering
    \scriptsize{(c) Ghost imaging, random masks, IXC}
    \end{minipage}%
    \begin{minipage}{0.15\linewidth}
    \centering
    \scriptsize{(d) Ghost imaging, orthogonal masks, XC}
    \end{minipage}%
    \begin{minipage}{0.15\linewidth}
    \centering
    \scriptsize{(e) Direct imaging}
    \end{minipage}\\
    
    \begin{minipage}{0.08\linewidth}
    \centering
    \scriptsize{(i) Noise free}
    \end{minipage}%
    \begin{minipage}{0.15\linewidth}
    \centering
    \scriptsize{~}\\
    \includegraphics[width=0.95\linewidth]{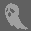}
    \end{minipage}%
    \begin{minipage}{0.15\linewidth}
    \centering
    \scriptsize{~}\\
    \includegraphics[width=0.95\linewidth]{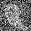}
    \end{minipage}%
    \begin{minipage}{0.15\linewidth}
    \centering
    \scriptsize{~}\\
    \includegraphics[width=0.95\linewidth]{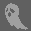}
    \end{minipage}%
    \begin{minipage}{0.15\linewidth}
    \centering
    \scriptsize{~}\\
    \includegraphics[width=0.95\linewidth]{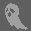}
    \end{minipage}%
    \begin{minipage}{0.15\linewidth}
    \centering
    \scriptsize{~}\\
    \includegraphics[width=0.95\linewidth]{./images/directImageNoiseFree.png}
    \end{minipage}\\

    \begin{minipage}{0.08\linewidth}
    \centering
    \scriptsize{(ii) Electronic read-out noise}
    \end{minipage}%
    \begin{minipage}{0.15\linewidth}
    \centering
    \scriptsize{SNR = 3.87 (3.90)}\\
    \includegraphics[width=0.95\linewidth]{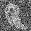}
    \end{minipage}%
    \begin{minipage}{0.15\linewidth}
    \centering
    \scriptsize{SNR = 2.85 (2.69)}\\
    \includegraphics[width=0.95\linewidth]{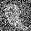}
    \end{minipage}%
    \begin{minipage}{0.15\linewidth}
    \centering
    \scriptsize{SNR = 27.7 (49.3)}\\
    \includegraphics[width=0.95\linewidth]{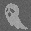}
    \end{minipage}%
    \begin{minipage}{0.15\linewidth}
    \centering
    \scriptsize{SNR = 61.2 (60.4)}\\
    \includegraphics[width=0.95\linewidth]{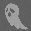}
    \end{minipage}%
    \begin{minipage}{0.15\linewidth}
    \centering
    \scriptsize{SNR = 1910 (1870)}\\
    \includegraphics[width=0.95\linewidth]{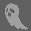}
    \end{minipage}\\
    
    \begin{minipage}{0.08\linewidth}
    \centering
    \scriptsize{(iii) Photon shot-noise}
    \end{minipage}%
    \begin{minipage}{0.15\linewidth}
    \centering
    \scriptsize{SNR = 3.87 (3.90)}\\
    \includegraphics[width=0.95\linewidth]{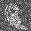}
    \end{minipage}%
    \begin{minipage}{0.15\linewidth}
    \centering
    \scriptsize{SNR = 2.06 (1.93)}\\
    \includegraphics[width=0.95\linewidth]{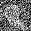}
    \end{minipage}%
    \begin{minipage}{0.15\linewidth}
    \centering
    \scriptsize{SNR = 2.42 (2.76)}\\
    \includegraphics[width=0.95\linewidth]{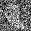}
    \end{minipage}%
    \begin{minipage}{0.15\linewidth}
    \centering
    \scriptsize{SNR = 2.74 (2.76)}\\
    \includegraphics[width=0.95\linewidth]{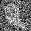}
    \end{minipage}%
    \begin{minipage}{0.15\linewidth}
    \centering
    \scriptsize{SNR = 88.0 (85.5)}\\
    \includegraphics[width=0.95\linewidth]{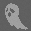}
    \end{minipage}

    \caption{
    Demonstration of the effects of (i) artefacts (noise-free), (ii) detector electronic read-out noise (modeled as Gaussian noise), and (iii) photon shot-noise (modeled as Poisson noise), for the following cases:
    (a) scanning probe imaging with pin-hole dimensions matching one pixel (i.e., $1/n^2$ mm$^2$); 
    (b) ghost imaging, using $J=2n^2$ random binary masks with $\mu_A = \sigma_A = 0.5$, recovered using scaled XC; 
    (c) ghost imaging as in (b) but recovered using $n^2$ Landweber iterations of IXC with $\alpha$ = 0.13, 0.016, 0.0020 for (a--c); 
    (d) ghost imaging using $J=n^2$ positions of an orthogonal FRT scanning mask (again with $\mu_A = \sigma_A = 0.5$), recovered using scaled XC (which is the inverse); and
    (e) direct imaging with an $n \times n$ pixel detector.
    The transmission image is a $1 \times 1$mm$^2$ binary stencil represented by a $32 \times 32$px array with black = 0.25, white = 0.75 ($\mu_T = 0.395$, $\sigma_T = 0.227$).
    Measured/recovered images are displayed with a grey-scale window of transmission in the range: [0,1.0). The flux was set so that (a--d) had the same total experiment time while (b--e) had the same dose incident on the stencil.
    $\mathcal{B} = 3.84 \times 10^5$ photons/mm$^2$/s with $t_0$ specified as follows:
    (a,d) 0.02s with 961 measurements,
    (b,c) 0.01s with 1922 measurements,
    (e) 9.61s.
    The SNR of the simulated results are provided where relevant (with theoretical values in brackets).
    \label{fig:gaussNoise}
    \label{fig:poissNoise}
    }
\end{figure*}

Several simulated demonstrations for this scenario have been presented in Fig.~\ref{fig:gaussNoise}.  In this figure, as well as for the remainder of the paper, we denote by XC the  cross-correlation method corresponding to using Eq.~(\ref{eq:adjoint}) and then correcting for the scale $\gamma$ in Eq.~(\ref{eq:gamma}); we denote by IXC the iterative cross-correlation method corresponding to the Landweber iteration in Eq.~(\ref{eq:landweber}).  A simulated scanning probe image is shown in Fig.~\ref{fig:gaussNoise}(a-ii), with the ghost image recovered from 1922 random masks by XC in Fig.~\ref{fig:gaussNoise}(b-ii) and IXC in Fig.~\ref{fig:gaussNoise}(c-ii).  The ghost image recovered by XC using 961 orthogonal masks (FRT based) is shown in Fig.~\ref{fig:gaussNoise}(d-ii). These all used the same total experiment time; both the random and orthogonal masks have $\sigma_A^2 = 0.25$ and the per-measurement noise had $\sigma_m=1.54$. From Eq.~(\ref{snrSpVsGiGauss}) we expect the SNR for (a-ii) to be reduced by a factor of 0.091 compared with (c-ii) [random masks] and reduced by 0.0645 compared with (d-ii) [orthogonal masks]. In the simulations we find ratios of 0.14 and 0.063 respectively. The result for orthogonal masks agrees with theory, however, the ratio is higher than expected for the random masks since the presence of noise has made it difficult for IXC to converge, causing the assumption that artefacts are negligible to become invalid. We see that despite this, SNR from ghost imaging remains higher than that for a scanning probe (or pin-hole mask).

\subsubsection{Comparison with direct imaging (constant dose)}
\label{sec:diVsGiGauss}

Here we consider a pixelated detector and assume that each pixel has the same characteristics as the bucket detector. Direct imaging has $n^2$ more flux than an $n \times n$ position scanning probe and $1/\mu_A$ more flux than ghost imaging. In this section we will maintain constant dose on the sample and compare the SNR of several schemes. Given a total dose of $\mathcal{P} \mu_A$ photons/px from a GI experiment, we estimate SNR for direct imaging (DI) as $\mathrm{SNR}_{DI} = \mathcal{P} \mu_A /  \sigma_m$. We again assume that degradation from image reconstruction artefacts is negligible, and that the SNR from ghost imaging is the reciprocal of Eq.~(\ref{eq:snrGauss}). As was the case previously, in order for this assumption to be valid (given no knowledge of the object) we require $J \ge n^2$. We can then determine the relationship:
\begin{equation}\label{snrDiVsGiGauss}
\mathrm{SNR}_{DI} = \sqrt{ \frac{ J \mu_A^2 }{ \sigma_A^2} } \textrm{SNR}_{GI}.
\end{equation}

When considering the mask property, $\mu_A^2/\sigma_A^2$, we note that the lower limit to this is defined by the binary case with $\mu_A^2/\sigma_A^2 = \mu_A / (1 - \mu_A)$. Therefore $\mu_A^2/\sigma_A^2 \ge 1 / (n^2 - 1)$ and $\mathrm{SNR}_{DI} \ge \textrm{SNR}_{GI}$ with equality given a pin-hole mask. We can conclude that in this scenario, in which one is given no knowledge of the object, (1) GI can do no better than direct imaging, and (2) typically $\mathrm{SNR}_{DI} \gg \textrm{SNR}_{GI}$.  The Fellgett advantage through increasing signal is lost when dose is fixed.

Several simulated demonstrations for this scenario have been presented in Fig.~\ref{fig:gaussNoise}.  The case of direct imaging is shown in Fig.~\ref{fig:gaussNoise}(e-ii).  The ghost image recovered from 1922 random masks using XC is shown in Fig.~\ref{fig:gaussNoise}(b-ii); the ghost image using IXC is in Fig.~\ref{fig:gaussNoise}(c-ii); the ghost image recovered by XC using 961 orthogonal masks (FRT based) is in Fig.~\ref{fig:gaussNoise}(d-ii). These all had the same total dose incident on the object; both the random and orthogonal masks have $\sigma_A^2 = 0.25$ and the per-measurement noise had $\sigma_m=1.54$. From Eq.~(\ref{snrDiVsGiGauss}) we expect the SNR for (e-ii) to be 38 times greater than (c-ii) [random masks] and 31 times greater than (d-ii) [orthogonal masks]. In the simulations we find ratios of 69 and 31 respectively. The result for orthogonal masks agrees with theory, however, the ratio is again higher than expected for the random masks, for the same reason already given in Sec.~\ref{sec:spVsGiGauss} above.

\subsection{Photon shot-noise}
\label{sec:poissonNoise}

For per-measurement read-out noise, the improvement in SNR through the Fellgett advantage \cite{fellgett1951theory} is proportional to the square-root of the number of channels that are multiplexed ($\mu_A n^2$ for GI). However, for photon shot-noise this advantage is lost since the noise level in each measurement also increases with the square-root of the number of channels that are multiplexed. We therefore do not expect GI to be advantageous in such a context.

\subsubsection{Comparison with scanning probe imaging (constant total exposure time, $\tau$)}
\label{sec:spVsGiPoisson}

We again assume the same hardware for both GI and a scanning probe and set the total experiment time to be $\tau$s. Scanning probe dwell time per pixel is therefore $\tau / n^2$s. Approximating the standard deviation of each measurement as the square root of the signal, (i.e., variance is $\sigma_p^2 \mathcal{P} \mu_T / n^2$) we then estimate SNR as $\textrm{SNR}_{SP} = \sqrt{ \mathcal{P} / \sigma_p^2 \mu_T n^2  }$. Here we are assuming that degradation from image reconstruction artefacts is negligible, and the SNR from ghost imaging reduces to the reciprocal of Eq.~(\ref{eq:snrPoisson}). We can then determine the relationship:
\begin{equation}\label{snrSpVsGiPoisson}
\textrm{SNR}_{SP} = \sqrt{  \frac{ \mu_A } { \sigma_A^2 } } \textrm{SNR}_{GI}.
\end{equation}

Observe that the smallest value that $\mu_A / \sigma_A^2$ can attain is determined by the binary case where $\mu_A / \sigma_A^2 = 1 / (1 - \mu_A)$. The absolute minimum corresponds to a pinhole mask (with $n \longrightarrow \infty$), i.e., a scanning probe. In this scenario, given no knowledge of the object, GI can achieve a reconstruction that is no better than that for a scanning probe, and is typically worse.

Several simulated demonstrations for this scenario have been presented in Fig.~\ref{fig:poissNoise}.  Scanning probe imaging is shown in Fig.~\ref{fig:poissNoise}(a-iii).  The ghost image recovered from 1922 random masks by XC is shown in Fig.~\ref{fig:poissNoise}(b-iii), with the corresponding IXC reconstruction in Fig.~\ref{fig:poissNoise}(c-iii).  The ghost image recovered by XC using 961 orthogonal masks (FRT based) is shown in Fig.~\ref{fig:poissNoise}(d-iii). These all used the same total experiment time.  Both the random and orthogonal masks have $\sqrt{ \mu_A / \sigma_A^2} = 1.4$, thus from Eq.~(\ref{snrSpVsGiPoisson}) we expect the SNR for (a-iii) to be 1.4 times higher than both (c-iii) [random masks] and (d-iii) [orthogonal masks]. In the simulations we find ratios of 1.6 and 1.4 respectively. The result for orthogonal masks agrees with theory, however, again we see the ratio is higher than expected for the random masks, for the same reason given in Sec.~\ref{sec:spVsGiGauss} above.

\subsubsection{Comparison with direct imaging (constant dose)}
\label{sec:diVsGiPoisson}

Given a pixelated detector we assume that each pixel has the same characteristics as the GI bucket detector. Here we keep dose on the sample constant and compare the SNR of several schemes. Direct imaging has $n^2$ more flux than an $n \times n$ position scanning probe and thus $1/\mu_A$ more flux than ghost imaging. Given a total dose of $\mathcal{P} \mu_A$ photons/px, SNR is estimated as $\textrm{SNR}_{DI} = \sqrt{  \mathcal{P} \mu_A / \sigma_p^2 \mu_T }$. Here again we are assuming that degradation from image reconstruction artefacts is negligible, and that the SNR from ghost imaging reduces to the reciprocal of Eq.~(\ref{eq:snrPoisson}). We can then determine the relationship:
\begin{equation}\label{snrDiVsGiPoisson}
\textrm{SNR}_{DI} = \sqrt{  \frac{ n^2 \mu_A^2 } { \sigma_A^2 } } \textrm{SNR}_{GI}.
\end{equation}
Noting again that $\sqrt{ \mu_A / \sigma_A^2} \ge 1$ and $n^2 \gg 1$, for this scenario (given no knowledge of the sample), direct imaging will always have a significantly higher SNR than ghost imaging.

Several simulated demonstrations for this scenario have been presented in Fig.~\ref{fig:poissNoise}.  Direct imaging is shown in Fig.~\ref{fig:poissNoise}(e-iii).  The ghost image recovered from 1922 random masks by XC is shown in Fig.~\ref{fig:poissNoise}(b-iii), with the corresponding IXC reconstruction in Fig.~\ref{fig:poissNoise}(c-iii).  The ghost image recovered by XC using 961 orthogonal masks (FRT based) is shown in Fig.~\ref{fig:poissNoise}(d-iii). These all had the same total dose incident on the object.  Both the random and orthogonal masks have $\sigma_A^2 = 0.25$. From Eq.~(\ref{snrDiVsGiGauss}) we expect the SNR for (e-iii) to be 31 times greater than (c-iii) [random masks] and (d-iii) [orthogonal masks]. In the simulations we find ratios of 36 and 32 respectively. The result for orthogonal masks agrees with theory, however, the ratio is again higher than expected for the random masks, for reasons given in Sec.~\ref{sec:spVsGiGauss}.

To summarize the findings in this section, given no knowledge of the sample:
\begin{itemize}
    \item A scanning probe can do no better than GI for per-measurement noise in the same experiment time since GI takes advantage of multiplexing (which increases signal while noise remains constant);
    \item GI can do no better than direct imaging (or a scanning probe) for per-measurement noise when subject to the same dose;
    \item GI can do no better than a scanning probe under shot-noise in the same experiment time;
    \item Direct imaging (or a scanning probe) will always be better than GI under shot-noise when subject to the same dose.
    \end{itemize}

It seems that the only way for GI to reduce dose is to capitalize on knowledge of the sample, enabling a low-artefact image to be reconstructed with very few measurements. This could be achieved through optimal mask design based on the {\it a priori} knowledge, or through reconstruction schemes that optimize the result subject to the {\it a priori} knowledge (e.g., compressed sensing, or maximum {\it a-posteriori} methods). In the next section, we will attempt to explain the experimental results from a recent 2D x-ray GI paper by Zhang {\it et al}.~\cite{zhang2018table} that seem to contradict the simulations and conclusions here.

\section{An explanation for contradictory results in the literature}
\label{sec:zhangExpl}

Experimental results indicating some extraordinary dose-reduction capabilities of x-ray GI were presented by Zhang {\it et al}.~\cite{zhang2018table}. This group used sandpaper to generate speckle in a repeatable manner, enabling a form of computational GI on a tabletop system. This paper represented the first demonstration of 2D x-ray GI, which is a remarkable achievement that marks a significant step forward in the field of x-ray GI \footnote{2D x-ray GI was also independently reported later in the same year, in the paper of Pelliccia {\it et al.}~\cite{pelliccia2018towards}.}. However, in this section we simulate their experiments and show that the associated results relating to dose, specifically the claim that use of the XC formula in Eq.~(\ref{eq:adjoint}) can yield ``a much higher contrast-to-noise ratio compared to projection x-ray imaging at the same low-radiation dose'' \cite{zhang2018table}, do not align with the predictions formulated in Sec.~\ref{sec:SNR} of our paper. Our simulation procedure, detailed in Appendix \ref{app:e}, attempted to replicate the experiments with a reasonable accuracy. The image quality from our simulations matches that of the high-dose experimental results reported in \cite{zhang2018table} reasonably well. However, a discrepancy arises between our simulations and their results when considering ultra-low dose experiments; the direct image measured appears to have lower contrast and be noisier than is the case for our simulations and the ghost image quality far exceeds that obtained in our simulations. We then propose a mechanism that explains this discrepancy and show that, when this suggested mechanism is included in our modeling, we are able to replicate both experimental results.

\subsection{Experiment simulation results}

Three images of the stencil described in Appendix \ref{app:e} are presented in Fig.~\ref{fig:simsSummary} and compared directly with the experiments of Ref.~\cite{zhang2018table}: (i) a ghost image reconstructed from $J=10,000$ with $t_0=150$ms (``Exp.~(i)''); (ii) a ghost image reconstructed from $J=10,000$ with $t_0=1\mu$s (``Exp.~(ii)''); (iii) a direct image recorded with $t_0 = 10$ms (``Exp.~(iii)''). Note that Exp.~(i) has the equivalent to total exposure, $Jt_0$, to that of Exp.~(ii). Results from the simulations are presented in Fig.~\ref{fig:simsSummary}(b) while those from Zhang {\it et al}.~are provided for reference in Fig.~\ref{fig:simsSummary}(a).

\begin{figure*}[ht!]
    \centering
    \begin{minipage}{0.28\linewidth}
    \centering
    \scriptsize{(a-i)}\\
    \includegraphics[width=0.9\linewidth]{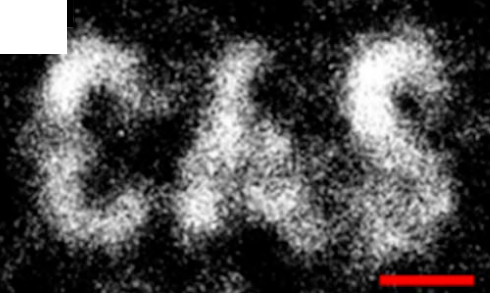}
    \end{minipage}%
    \begin{minipage}{0.28\linewidth}
    \centering
    \scriptsize{(b-i)}\\
    \includegraphics[width=0.9\linewidth]{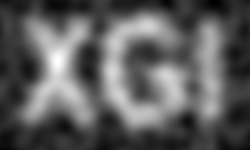}
    \end{minipage}%
    \begin{minipage}{0.28\linewidth}
    \centering
    \scriptsize{(c-i)}\\
    \includegraphics[width=0.9\linewidth]{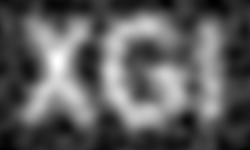}
    \end{minipage}\\
    \begin{minipage}{0.28\linewidth}
    \centering
    \scriptsize{(a-ii)}\\
    \includegraphics[width=0.9\linewidth]{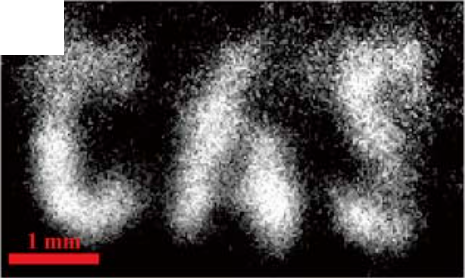}
    \end{minipage}%
    \begin{minipage}{0.28\linewidth}
    \centering
    \scriptsize{(b-ii)}\\
    \includegraphics[width=0.9\linewidth]{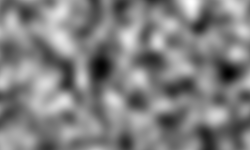}
    \end{minipage}%
    \begin{minipage}{0.28\linewidth}
    \centering
    \scriptsize{(c-ii)}\\
    \includegraphics[width=0.9\linewidth]{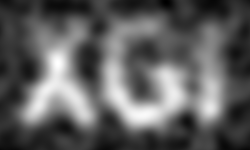}
    \end{minipage}\\
    \begin{minipage}{0.28\linewidth}
    \centering
    \scriptsize{(a-iii)}\\
    \includegraphics[width=0.9\linewidth]{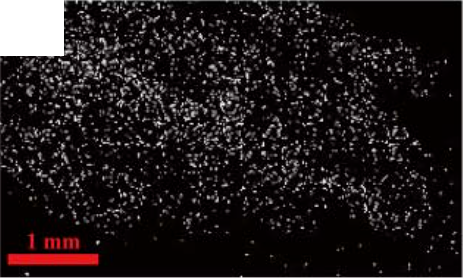}
    \end{minipage}%
    \begin{minipage}{0.28\linewidth}
    \centering
    \scriptsize{(b-iii)}\\
    \includegraphics[width=0.9\linewidth]{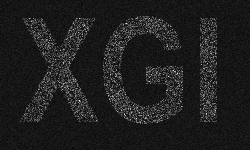}
    \end{minipage}%
    \begin{minipage}{0.28\linewidth}
    \centering
    \scriptsize{(c-iii)}\\
    \includegraphics[width=0.9\linewidth]{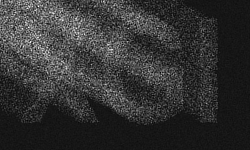}
    \end{minipage}
    \caption{The resulting images from experiments (i-iii) in Ref.~\cite{zhang2018table}:  row (i) corresponds to ghost imaging with $J=10,000$ and $t_0=150$ms, row (ii) corresponds to ghost imaging with $J=10,000$ and $t_0=1\mu$s, and row (iii) corresponds to direct imaging with $t_0 = 10$ms. Column (a) shows the results from Zhang et al. with permission from \cite{zhang2018table} \copyright The Optical Society; column (b) depicts the corresponding simulations as described in Appendix \ref{app:e}; column (c) presents the simulations revisited with CCD read-out in the absence of a shutter (as described in Sec.~\ref{sec:simsCcdReadout}).}
    \label{fig:simsSummary}
\end{figure*}

\begin{figure*}[ht!]
    \centering
    \begin{minipage}{0.28\linewidth}
    \centering
    \scriptsize{(a)}\\
    \includegraphics[width=0.9\linewidth]{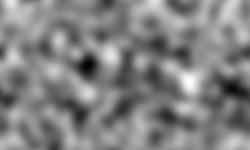}
    \end{minipage}%
    \begin{minipage}{0.28\linewidth}
    \centering
    \scriptsize{(b)}\\
    \includegraphics[width=0.9\linewidth]{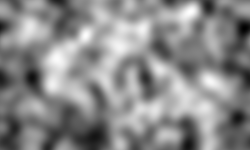}
    \end{minipage}%
    \begin{minipage}{0.28\linewidth}
    \centering
    \scriptsize{(c)}\\
    \includegraphics[width=0.9\linewidth]{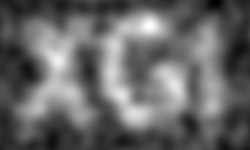}
    \end{minipage}
    \caption{The ghost image recovered from 10,000 bucket measurements with (a) 10$\mu$s, (b) 100$\mu$s, and (c) 1ms exposure per measurement. The CCD is assumed to be shielded from the continuous-wave x-ray source during CCD read-out. The field-of-view is 5mm across.}
    \label{fig:simsNoShutter}
\end{figure*}

Observe that for high dose, i.e., Exp.~(i) where $Jt_0 = 1500$s, the x-ray GI (XGI) results are of comparable quality. However, for the two ultra-low dose cases, Exp.~(ii-iii) where total experiment time $\tau = Jt_0 = 0.01$s, the results disagree markedly. The XGI simulation in Fig.~\ref{fig:simsSummary}(b-ii) has produced only noise. Figure \ref{fig:simsNoShutter} shows that in order to reconstruct something meaningful, given $J=10,000$ speckle patterns, $t_0$ must be increased 1000 fold, to on the order of 1ms. The conclusion drawn in Ref.~\cite{zhang2018table} from the experimental results is that ``for a given $Jt_0$, XGI is better than projection imaging,'' however, our simulation results suggest that the opposite is true. The ultra-low dose direct image measured is worse than simulations predict while the ultra-low dose ghost image quality far exceeds that expected from simulations. 

We believe that the mechanism that can explain both the degradation of ultra-low dose direct imaging and the improvement of ultra-low dose GI is that no shutter was used and the detector continues to collect signal during the charge-coupled device (CCD) read-out time (which is on the order of 100ms). It seems that the authors in Ref.~\cite{zhang2018table} acknowledged the possibility of this problem with the statement: ``Of course, in practice we would need to shutter the beam before the object or use a pulsed source.'' In the remainder of this section we will first describe, then demonstrate experimentally, the effect of performing CCD read-out in the presence of a continuous wave (CW -- as opposed to pulsed) x-ray source. We will then simulate the experiments performed by Zhang {\it et al}.~\cite{zhang2018table} without a shutter.

\subsection{Simulation results incorporating CCD read-out time}
\label{sec:simsCcdReadout}

\begin{figure*}[ht!]
    \centering
    \begin{minipage}{0.28\linewidth}
    \centering
    \scriptsize{(a)}\\
    \includegraphics[width=0.9\linewidth]{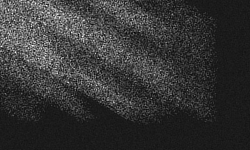}
    \end{minipage}%
    \begin{minipage}{0.28\linewidth}
    \centering
    \scriptsize{(b)}\\
    \includegraphics[width=0.9\linewidth]{./images/directImaging10msNoShutter.png}
    \end{minipage}%
    \begin{minipage}{0.28\linewidth}
    \centering
    \scriptsize{(c)}\\
    \includegraphics[width=0.9\linewidth]{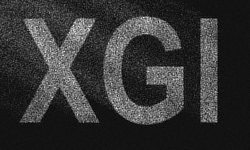}
    \end{minipage}
    \caption{Direct images (with first speckle mask in place) simulated assuming exposure times of (a) 1ms, (b) 10ms, and (c) 100ms. The images simulate the result of CCD read-out in 0.93s with no shutter in place. Note that all images assumed the camera to have been ``cleared'' before exposure.  The diagonal smearing, which is particularly evident in panels (a) and (b), gives a clear signature that the detector continues to collect signal during the CCD read-out time. The field-of-view is 5mm across.}
    \label{fig:simsDirectImageNoShutter}
\end{figure*}

A description and demonstration of the smearing effect resulting from x-ray exposure during CCD read-out time is presented in Appendix \ref{app:f}. Here we have incorporated a ``CCD read-out'' function that replicates that process. We have used $t_1 = 0.93$s for direct imaging and $t_1 = 0.12$s for ghost imaging (since it was binned $8 \times 8$ for faster readout). The CCD has 1300 rows, therefore the dwell time per row read-out was $t_1/1300$ = 720$\mu$s for direct imaging and 92$\mu$s for ghost imaging. We observe from Fig.~\ref{fig:simsSummary}(a-iii) (and from correspondence with the authors of Ref.~\cite{zhang2018table}) that the stencil was rotated approximately 60$^\circ$ relative to the CCD. Figure \ref{fig:simsDirectImageNoShutter} presents the results of incorporating both the CCD read-out time and 60$^\circ$ rotation for $t_0 = 1$ms, 10ms, and 100ms. We are able to replicate the appearance of Fig.~\ref{fig:simsSummary}(a-iii) and we observe that an exposure time as low as $t_0 = t_1 / 10$ is sufficient for the diagonal smearing, associated with CCD read-out time,  to become negligible.  

The results from repeating the simulations of Exp.~(i-iii), this time incorporating CCD read-out in the absence of a shutter, are presented in Fig.~\ref{fig:simsSummary}(c). The ultra-low dose cases, Exp.~(ii-iii), now more closely resemble those from the experiments in Ref.~\cite{zhang2018table}. The absence of a shutter is detrimental to the direct image since the ``read-out smear'' obscures the primary image. In ghost imaging the data from the CCD is binned to a single ``bucket'' measurement. Therefore the absence of a shutter is beneficial as the additional x-ray exposure during read-out is as valuable as the initial x-ray exposure. Therefore we could say that the equivalent exposure in the presence of a shutter for Exp.~(ii) is $J(t_0+t_1) = 1200$s. The corresponding dose is a factor of $10^5$ times larger than that which would be anticipated from the nominal exposure time alone,   approaching the high dose case in Exp.~(i), which is 1500s with a shutter (or 2700s in the absence of a shutter).

Based on correspondence with the authors of Ref.~\cite{zhang2018table} we understand that this effect was mitigated to some degree by: (1) removing the ``smear'' above the stenciled letters before summing to a ``bucket'' value; and (2) subtracting an estimate of the smear magnitude based on signal in the darkfield region. The letters are approximately 125 pixels high, therefore, considering (1), the effective readout time could be reduced to 0.012s such that $J(t_0+t_1) = 120$s. However, this is still a $10^4$-fold larger dose than anticipated.

\section{Discussion}
\label{sec:disc}

We hope the present paper clarifies the circumstances under which GI may be a useful tool to employ, and under what circumstances other tools (e.g.~those based on conventional imaging) are more appropriate.  As GI transitions beyond proof-of-concept foundational studies, practical considerations such as dose reduction are likely to receive increased attention.  A route, from the now-established proof-of-concept to a clearer practical understanding of the circumstances in which GI is actually advantageous, is the core motivation for undertaking the work presented here.

Leveraging suitable {\it a priori} knowledge is a necessary (but not sufficient) condition for classical ghost imaging to achieve dose reduction relative to conventional imaging employing position-sensitive detection.  Both mask design and data processing can utilize such {\it a priori} knowledge in a GI setting.  It would be interesting to further investigate both strategies, in the specific context of dose reduction. One could envisage specially designed ghost-imaging masks tailor-made for answering particular questions regarding particular classes of sample, e.g.~breast-cancer screening for persons of a known gender and known age-range, security screening of passenger luggage, defect detection in particular manufactured products etc.  At the post-detection data-analysis stage, relevant prior knowledge could be readily incorporated into a plethora of data-analysis approaches (including but not limited to those that incorporate machine learning) with a view to achieving GI-enabled dose reduction.

It is also worth pointing out that GI ``does not need to be about images''.  For example, one may want to use an ensemble of GI-bucket measurements to reconstruct a single parameter of interest regarding a sample.  Such sample-parameters may be binary, e.g.~a yes/no diagnosis for a patient or a reject/retain decision for industrial-sample quality control.  These sample-parameters may also be real numbers, such as porosity, mean curvature, neck-to-void ratio, surface-area-to-volume-ratio, connected-volume-to-disconnected-volume ratio, Euler-number density, mean chord length, fractal dimension, dislocation density, defect density etc.~of samples such as sponges, foams, glasses, alloys, chemical-reaction catalysts, oil-bearing rocks, lung tissue etc.  Questions regarding dose-reduction could be investigated in this context where one does not seek to reconstruct an image of the sample, but rather to reconstruct one or more parameters regarding that sample.  A conventional-imaging approach would measure one or more images, and subsequently process these images to extract the desired parameter or parameters; GI, on the other hand, never measures image data in the first place, and could process the bucket signals directly so as to yield the desired parameter(s)  without going via the intermediate step of reconstructing an image.  The question of dose-reduction and SNR could be revisited in light of the reconstruction of sample parameters rather than sample images.  Since this is a different question to that of optimized {\em image} SNR, we expect there to be different answers; perhaps there will be non-imaging parameter-measurement situations in which GI gives superior SNR, that have not been considered in the present paper.

One can also consider questions complementary to the above reduced-dimensionality settings: those using GI to obtain higher-dimensional reconstructions, e.g.~for three spatial dimensions (3D) via ghost tomography \cite{KingstonOptica2018, KingstonIEEE2019}, 4D (e.g.~three spatial dimensions and one time dimension or two spatial dimensions plus one energy dimension and one time dimension), 5D (e.g.~three spatial dimensions, one energy or material-composition dimension and one time dimension), etc.  ``Conventional imaging'' in many of these contexts become computational methods when a probe to scan individual volume elements does not exist. It would be interesting to explore the potential for classical GI to achieve dose-reduction in these higher-dimensional contexts, both with and without the leveraging of suitable prior knowledge.

GI may still be important in contexts where conventional imaging, while superior in principle, is not available or practical. Bucket detectors used for GI are much smaller, cheaper, and will often possess higher energy resolution and/or temporal resolution than corresponding position-sensitive detectors. Dose-reduction is one part of a broader suite of tradeoffs that include financial cost, practicality, availability, simplicity, detector weight, detector volume, experiment miniaturization etc.  Even when GI is no better than conventional imaging from the point of view of dose/SNR, the findings of our paper may be useful within this broader suite of tradeoffs.

\section{Conclusions}
\label{sec:conc}

We have examined the inherent signal-to-noise ratio (SNR) properties of classical ghost imaging (GI) as a function of imaging parameters such as: number of masks; mean mask transmission; mask transmission variance; exposure time. By ``inherent'' we mean that no {\it a priori} knowledge of the object being imaged was exploited.
The SNR under Poisson and per-measurement Gaussian noise models was explored and compared with conventional imaging techniques (direct and scanning probe imaging) under conditions of constant dose and constant experiment time. 
It was found that: (i) a scanning probe can do no better than GI for per-measurement noise in the same experiment time; (ii) GI can do no better than conventional imaging for per-measurement noise when subject to the same dose; (iii) 
GI can do no better than a scanning probe under shot-noise in the same experiment time; (iv) conventional imaging will always be better than GI under shot-noise when subject to the same dose.

An explanation was proposed for the experimental results from a recent 2D x-ray GI paper by Zhang {\it et al}.~\cite{zhang2018table} that seem to contradict these conclusions. The use of a CCD without a shutter under continuous wave x-ray illumination both improves GI results and degrades conventional imaging causing GI to appear to give better SNR when given the same dose.

We conclude that for GI to reduce dose it must capitalize on {\it a priori} knowledge of the sample to yield low-artefact images reconstructed from very few measurements. This could be achieved through optimal mask design based on the {\it a priori} knowledge, or through reconstruction schemes that optimize the result subject to the {\it a priori} knowledge (e.g., compressed sensing, or maximum {\it a-posteriori} methods). While work in compressed sensing literature directs how to reduce the number of measurements, we wish to emphasize that {\it reducing the number of measurements} is not synonymous with {\it reducing dose}; it is not a trivial undertaking. A first step towards this goal could involve understanding how to modify masks, given {\it a priori} knowledge, in order to reduce dose compared with GI using the original masks. A second step would then be understanding the limits of such a process and under what conditions the result can improve on conventional imaging.

\begin{acknowledgments}
We gratefully acknowledge useful correspondence with Timur E. Gureyev, Alexander Rack, Thomas J. Lane, Daniel Ratner, and with the authors of Ref.~\cite{zhang2018table}. AMK acknowledges the financial support of the Australian Research Council Industrial Transformation Training Centre IC180100008.
\end{acknowledgments}

\bibliography{aps-main}

\appendix

\section{Traditional ghost image reconstruction is adjoint to cross-correlation}
\label{app:a}

Recall that given $J$ patterned illuminations, $A = \{A_1(x,y), A_2(x,y), ..., A_J(x,y)\}$, where $x$ and $y$ are detector coordinates, the set of bucket signals measured, $b = \{b_1, b_2, ..., b_J\}$, is modeled as a set of cross-correlations of $A$ with the transmission image of the object of interest, i.e., Eq.~(\ref{eq:xc}). The traditional ghost image reconstruction equation presented in the literature, typically without derivation, is the following \cite{katz2009compressive,Bromberg2009ghost}:
\begin{equation*}
    \widehat{T}(x,y) = \sum_j  A_j(x,y) \tilde{b}_j,
\end{equation*}
where $\tilde{b}_j = b_j - \langle b \rangle$ are mean-corrected bucket signals with $\langle b \rangle$ denoting the mean bucket signal, i.e., $\langle b \rangle = (1/J)\sum_j b_j$. Observe that this reconstruction operation acts on $\tilde{b}_j$ rather than $b_j$. Let us define a mean-corrected form of the cross-correlation (Eq.~(\ref{eq:xc})) that results in $\tilde{b}_j$ as follows:
\begin{equation*}
    \tilde{b}_j = \sum_x \sum_y \tilde{A}_j(x,y) T(x,y) = [ \tilde{\mathcal{A}}T ]_j,
\end{equation*}
where $\tilde{A}_j(x,y) = A_j(x,y) - \langle A(x,y) \rangle$ are mean-corrected illuminating speckle patterns with $\langle A(x,y) \rangle$ denoting the mean speckle intensity at position $(x,y)$, i.e., $\langle A(x,y) \rangle = (1/J)\sum_j A(x,y)$. Traditional ghost image reconstruction (Eq.~(\ref{eq:adjoint})) can be derived as the adjoint of this operator, i.e., $\widehat{T} = \tilde{\mathcal{A}}^\dagger \tilde{b}$. The adjoint (Hermitian conjugate) of operator $\tilde{\mathcal{A}}$ can be found through the inner product identity, $\left< \tilde{\mathcal{A}}T, \tilde{b} \right > = \langle T , \tilde{\mathcal{A}}^\dagger \tilde{b} \rangle$, as follows:
\begin{eqnarray*}
\left< \tilde{\mathcal{A}}T, \tilde{b} \right > & = & \sum_j \left( \sum_x \sum_y \tilde{ A}_j(x,y) T(x,y)  \right) \tilde{b}_j \nonumber \\
& = & \sum_j \left( \sum_x \sum_y \left( A_j(x,y) - \langle A(x,y) \rangle \right) T(x,y)  \right) \tilde{b}_j \nonumber \\
& = & \sum_x \sum_y T(x,y) \sum_j \left ( A_j(x,y) - \langle A(x,y) \rangle \right) \tilde{b}_j \nonumber \\
& = & \sum_x \sum_y T(x,y) \nonumber \\
&& \qquad \left ( \sum_j A_j(x,y) \tilde{b}_j - \langle A(x,y) \rangle \sum_j  \tilde{b}_j \right)\nonumber \\
& = & \sum_x \sum_y T(x,y) \sum_j A_j(x,y) \tilde{b}_j \nonumber \\
& = & \langle T , \tilde{\mathcal{A}}^\dagger \tilde{b}_j\rangle.
\end{eqnarray*}

\noindent In general, the linear operator $\mathcal{A}$ will not be such that its adjoint coincides with its inverse.  By definition, $\mathcal{A}^\dagger=\mathcal{A}^{-1}$ if and only if $\mathcal{A}$ is unitary.  Unitarity holds if the illumination patterns form an orthogonal basis.

\section{Calculation of ghost imaging normalization scale, $\gamma^{-1}$}
\label{app:b}

For a set of independent random patterned illuminations (see e.g.~the  example of patterned illumination and GI results depicted in Figs.~\ref{fig:PSF}(a--b) and \ref{fig:giPsfDemo}(a--b)) we assume spatial statistical stationarity and observe the following property:
\begin{equation*} 
    \sum_j \tilde{A}_j(x',y')\tilde{A}_j(x,y) \approx J \sigma_A^2 \delta(x-x',y-y').
\end{equation*}
Inserting this property into Eq.~(\ref{eq:gamma}) we arrive at:
\begin{eqnarray*}
    \gamma
    & = & \sum_x \sum_y \frac{1}{n^2} \sum_j [\tilde{A}_j \circ \tilde{A}_j] (x,y)  \nonumber\\
    & = & \sum_x \sum_y \frac{1}{n^2} \sum_j \sum_{x^*} \sum_{y^*} \tilde{A}_j(x^*,y^*)\tilde{A}_j(x^* + x, y^* + y) \nonumber\\
    & \approx & \frac{1}{n^2} \sum_{x^*} \sum_{y^*} J \sigma_A^2 \sum_x \sum_y \delta(x,y) \nonumber\\
    & = & J \sigma_A^2.
\end{eqnarray*}

Consider the case where the resolution of the random speckle images is reduced, e.g.,  due to factors such as penumbral blurring from a non-point source, detector cross-talk, etc. Let this be modeled as a blur through convolution with a 2D Gaussian kernel with standard deviation $\sigma_{g}$, i.e.,
\begin{eqnarray*}
    A_j^*(x,y)
    & = & \{ A_j \ast K_{\sigma_g} \} (x,y) \nonumber \\
    & = & \sum_x' \sum_y' A_j(x',y')K_{\sigma_g}(x-x',y-y'), \nonumber
\end{eqnarray*}
where:
\begin{equation*}
  K_{\sigma_g}(x,y) = \frac{1}{2 \pi \sigma_g^2}\exp{\left[ \frac{-(x^2 + y^2)}{2\sigma_g^2} \right] }.
\end{equation*}
Example patterned illumination and GI results for this scenario are depicted in Fig.~\ref{fig:PSF}(c--d) and Fig.~\ref{fig:giPsfDemo}(c--d). The variance of these blurred masks becomes $\sigma_{A^*}^2 = \sigma_A^2 / 4 \pi \sigma_g^2$. Noting that the result of the convolution of two Gaussian distributions, denoted $K_{\sigma_1} \ast K_{\sigma_2}$, is also a Gaussian distribution with $\sigma^2 = \sigma_1^2 + \sigma_2^2$, and again assuming spatial statistical stationarity we observe the following property:
\begin{eqnarray*}
    & \sum_j & \tilde{A}_j^*(x',y')\tilde{A}_j^*(x,y) \\
    & \approx & J \sigma_{A}^2 \frac{1}{4 \pi \sigma_g^2} \exp{\left\{ \frac{-[(x'-x)^2 + (y'-y)^2]}{4\sigma_g^2} \right\}} \\
    & = & J \sigma_{A^*}^2 \exp{\left\{ \frac{-[(x'-x)^2 + (y'-y)^2]}{4\sigma_g^2} \right\}}.
\end{eqnarray*}
Inserting this property into (\ref{eq:gamma}) we arrive at:
\begin{eqnarray*}
    \gamma
    & = & \sum_x \sum_y \frac{1}{n^2} \sum_j \sum_{x^*} \sum_{y^*} \tilde{A}_j(x^*,y^*)\tilde{A}_j(x^* + x, y^* + y) \nonumber\\
    & \approx & \frac{1}{n^2} \sum_{x^*} \sum_{y^*} J  \sigma_{A^*}^2 \sum_x \sum_y \exp{\left[ \frac{-(x^2 + y^2)}{4\sigma_g^2} \right]} \nonumber\\
    & \approx & \frac{1}{n^2} \sum_{x^*} \sum_{y^*} J  \sigma_{A^*}^2  4 \pi \sigma_g^2 \nonumber\\
    & = & 4 \pi J \sigma_g^2 \sigma_{A^*}^2.
\end{eqnarray*}

\section{Derivation of SNR as functions of imaging parameters}
\label{app:c}

\paragraph{Noise-free case with random masks:} Consider the high-photon-flux (or noise-free) limit, given a set of $J$ independent, random masks, $A_j(x,y)$, with a mean value of $\mu_A$ and a standard deviation of $\sigma_A$. The variance of the bucket signals, $b_j$, determined from Eq.~(\ref{eq:xc}) is:
\begin{equation*}
    \sigma^2_b \approx n^2 (\mu^2_T + \sigma^2_T) \sigma^2_{A},
\end{equation*}
where $\mu_T$ and $\sigma_T$ are the mean and standard deviation of object transmission respectively. The simplest situation to consider is where $T(x,y)$ is constant; the variance (or mean square-error, $\epsilon^2_{\widehat{T}}$) in ghost image reconstruction using the adjoint (Eq.~(\ref{eq:adjoint})) is found as:
\begin{eqnarray}
    \epsilon^2_{\widehat{T}} = \sigma^2_{\widehat{T}}
    & \approx & J \sigma^2_A \sigma^2_b \nonumber \\
    & \approx & J \sigma^4_A n^2 (\mu^2_T + \sigma^2_T).
\end{eqnarray}
The standard deviation (or RMSE) of GI is then the square root of this, scaled by $\gamma^{-1} = 1/J\sigma^2_A$. Insert this into the definition of SNR to arrive at:
\begin{equation*}
    \mbox{SNR}_0(T,\widehat{T}) \approx \sqrt{ \frac{ J \mu_T} { n^2 (\mu^2_T + \sigma^2_T) } }.
\end{equation*}

\paragraph{Noise-free case with orthogonal scanning masks:} Again consider the high-photon-flux (or noise-free) limit, this time given a scanning mask that is orthogonal under translation (such as a uniformly redundant array \cite{Gottesman1989newFamily} or a mask that is generated through the finite Radon transform \cite{cavy2015construction}).  In this case, SNR analysis is performed from a different perspective. Here at most $J=n^2$ masks are utilized and error only arises when $J<n^2$. We consider the $(n^2-J)$ unmeasured bucket signals to be present but with a zero signal. The square-error in each ``missing'' bucket signal is then estimated to be $\sigma^2_b \approx n^2 \sigma^2_T \sigma^2_A$. Note that in this case $\mu_T$ disappears since the masks are strictly constant-mean, i.e., $\sum_x \sum_y A_j(x,y) = k ~ \forall ~ j \in [1,J]$. The square-error in the reconstructed image using the adjoint (Eq.~(\ref{eq:adjoint})) is found as:
\begin{equation*}
    \epsilon^2_{\widehat{T}} \approx (n^2-J) \sigma^2_A \sigma^2_b \approx (n^2-J) n^2 \sigma^4_A \sigma^2_T.
\end{equation*}
The RMSE is the square root of this scaled by $\gamma^{-1} = 1/n^2\sigma^2_A$ (here we have assumed $n^2$ measurements were taken as described above). Insert this into the SNR definition to arrive at:
\begin{equation*}
    \textrm{SNR}^\perp_0(\widehat{T},T) \approx \sqrt{\frac{ n^2} { (n^2 - J) \sigma^2_T} }.
\end{equation*}

\paragraph{Photon shot-noise:} A Poisson distribution scaled by $\sigma_p$ is employed to model photon shot-noise. In the limit where there are no artefacts from GI, (i.e., $J \gg n^2$ for the random mask case or $J=n^2$ for orthogonal masks), SAR $\rightarrow \infty$ and SNR approaches a more ``conventional'' imaging definition. The mean square-error (MSE), $\epsilon^2_b$, in each bucket signal is proportional to the mean number of photons detected:
\begin{equation*}
    \epsilon^2_b \approx \sigma^2_p \left( \frac{ \mathcal{P} }{ J } \mu_A  n^2 \mu_T \right),
\end{equation*}
where $\sigma^2_p$ is a constant of proportionality. The MSE (or variance) in the reconstructed image using the adjoint (Eq.~(\ref{eq:adjoint})) is found as:
\begin{equation*}
    \epsilon^2_{\widehat{T}} \approx J \sigma^2_A \epsilon^2_b \approx \mathcal{P} \sigma^4_A \sigma^2_p n^2 \mu_A \mu_T.
\end{equation*}
The RMSE is the square root of this scaled by $\gamma^{-1} = 1 / \mathcal{P} J \sigma^2_A$, i.e.,
\begin{equation*}
\textrm{RMSE}_p(\widehat{T},T) = \sqrt{ \frac{ \sigma_p^2 \mu_A \mu_T n^2 }{ \mathcal{P} \sigma_A^2 } },
\end{equation*}
where $\mu_A$ is the mean transmission of the speckle masks.

\paragraph{Per-measurement electronic read-out noise:} A Gaussian distribution with uniform standard deviation, $\sigma_m$, is used to model per-measurement electronic read-out noise. In the same limit as for the shot-noise case, the MSE in each bucket signal is simply $\epsilon^2_b = \sigma^2_m$. The MSE (or variance) in the reconstructed image using the adjoint (Eq.~(\ref{eq:adjoint})) is found as:
\begin{equation*}
    \epsilon^2_{\widehat{T}} \approx J \sigma^2_A \epsilon^2_b = J \sigma^2_A \sigma^2_m.
\end{equation*}
The RMSE is the square root of this scaled by $\gamma^{-1} = 1 / \mathcal{P} J \sigma^2_A$, i.e.,
\begin{equation*}
\mathrm{RMSE}_m(\widehat{T},T) = \sqrt{ \frac{ J \sigma_m^2 }{ \mathcal{P}^2 \sigma_A^2 } }.
\end{equation*}

\section{Demonstration of SNR response to imaging parameters}
\label{app:d}

We performed a suite of simulations to demonstrate the effect of each imaging parameter on the recovered ghost image SNR. For the simulations, a suite of  $n \times n = 64 \times 64$ px uniformly random arrays, with $T(x,y) \in [0.0,1.0]$, was used as the 1mm$^2$ transmission image of the object.  The values were scaled and offset to modify the image mean and standard deviation. The pixel-pitch is therefore $1/n$mm and the pixel area is $1/n^2$mm$^2$. A parallel x-ray beam is assumed, with a set of 1mm$^2$ speckle masks having a resolution of $2/n$mm creating the patterned illuminations for ghost imaging. Given an incident photon flux of $\mathcal{B} = 4.1 \times 10^5$ photons/s/mm$^2$ and an exposure time of $t_0 = 0.01$s per bucket measurement, i.e., $\Xi = Jt_0/n^2$s/px, the total experiment time is then $\tau = J t_0$ seconds. The resulting speckled illuminations are represented by $n \times n$ pixel random binary images with $\mathcal{B} t_0 A_j(x,y) / n^2$ photons per pixel for $j \in [1,J]$.

\begin{figure}[ht!]
    \centering
    \begin{minipage}{0.5\linewidth}
    \centering
    \scriptsize{(a)}\\
    \includegraphics[width=\linewidth]{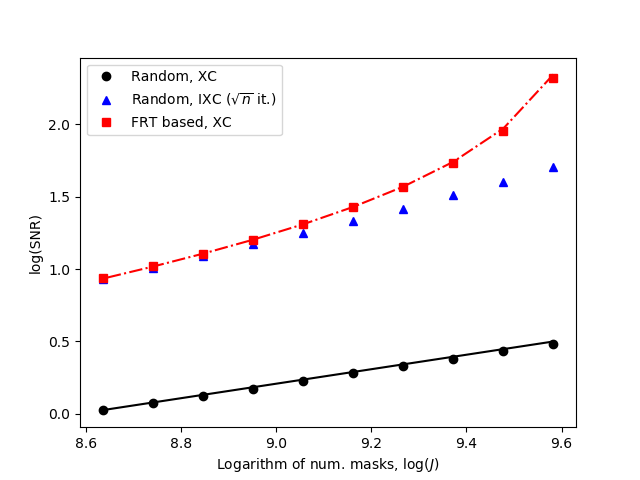}
    \end{minipage}%
    \begin{minipage}{0.5\linewidth}
    \centering
    \scriptsize{(b)}\\
    \includegraphics[width=\linewidth]{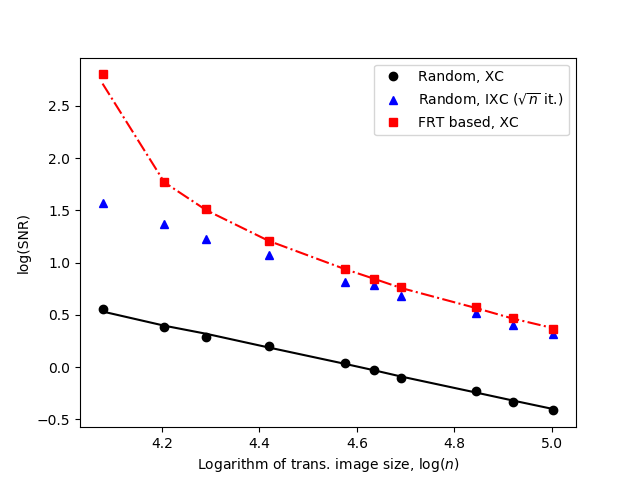}
    \end{minipage}\\
    
    \begin{minipage}{0.5\linewidth}
    \centering
    \scriptsize{(c)}\\
    \includegraphics[width=\linewidth]{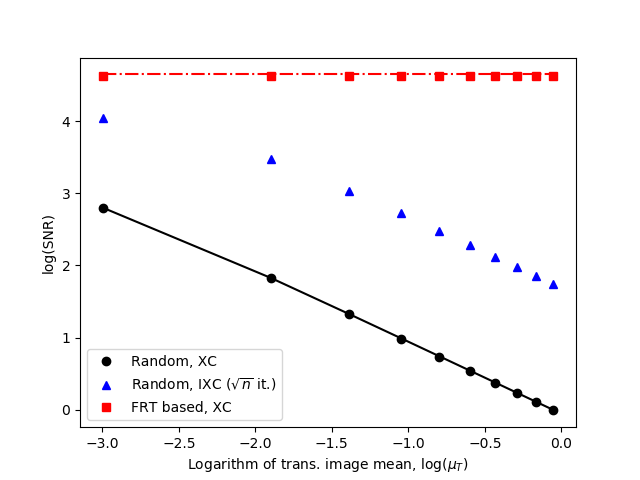}
    \end{minipage}%
    \begin{minipage}{0.5\linewidth}
    \centering
    \scriptsize{(d)}\\
    \includegraphics[width=\linewidth]{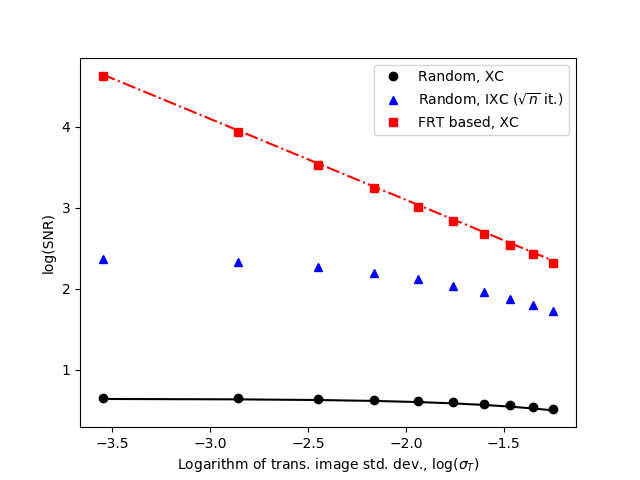}
    \end{minipage}
    \caption{Demonstration of the SNR of ghost imaging for random and orthogonal masks as a function of imaging parameters: (a) number of masks, $J$; (b) image dimension, $n$; (c) mean object transmission, $\mu_T$; (d) standard deviation of object transmission, $\sigma_T$. Markers in the plot are simulation results while the lines connect the expected result according to Eqs.~(\ref{eq:snrNoiseFreeRandom}) and (\ref{eq:snrNoiseFreeOrtho}) (for random and orthogonal masks respectively). Ghost images recovered using iterative cross-correlation (IXC) (Eq.~(\ref{eq:landweber})) have employed $\sqrt{n}$ iterations with regularization term $\alpha = 0.5$.}
    \label{fig:snrStudyNoiseFreePlots}
\end{figure}

\paragraph{Noise-free case:} Simulations demonstrating SNR$_0$ (Eq.~(\ref{eq:snrNoiseFreeRandom})) and SNR$^\perp_0$ (Eq.~(\ref{eq:snrNoiseFreeOrtho})) as a function of imaging parameters $J,n,\mu_T,\sigma_T$ are presented in Fig.~\ref{fig:snrStudyNoiseFreePlots} under the high-photon-flux (or noise-free) limit for the cases of an orthogonal scanning mask, as well as random masks using both the scaled adjoint (Eq.~(\ref{eq:adjoint})) and Landweber iteration (Eq.~(\ref{eq:landweber})) for image reconstruction. Equations~(\ref{eq:snrNoiseFreeRandom}) and (\ref{eq:snrNoiseFreeOrtho})   have been overlaid to demonstrate their consistency with these simulations.

\begin{figure}[ht!]
    \centering
    \begin{minipage}{0.5\linewidth}
    \centering
    \scriptsize{(a)}\\
    \includegraphics[width=\linewidth]{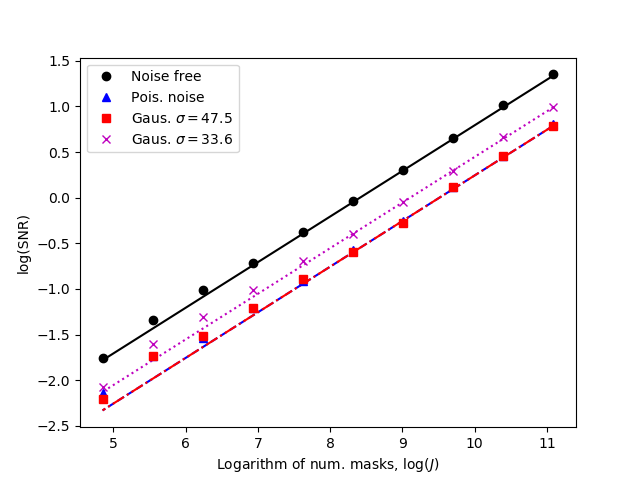}
    \end{minipage}%
    \begin{minipage}{0.5\linewidth}
    \centering
    \scriptsize{(b)}\\
    \includegraphics[width=\linewidth]{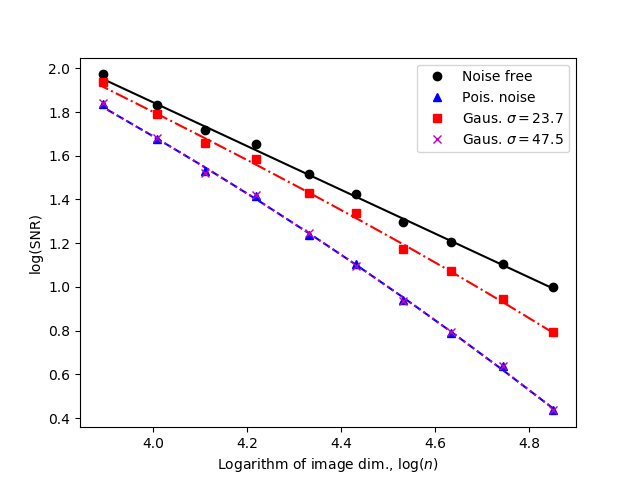}
    \end{minipage}\\
    
    \begin{minipage}{0.5\linewidth}
    \centering
    \scriptsize{(c)}\\
    \includegraphics[width=\linewidth]{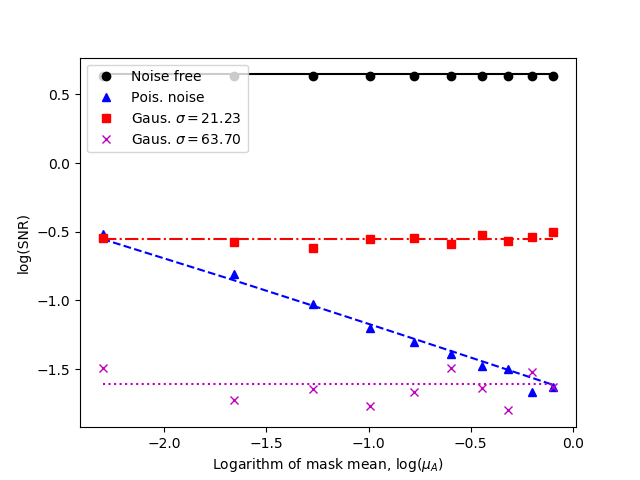}
    \end{minipage}%
    \begin{minipage}{0.5\linewidth}
    \centering
    \scriptsize{(d)}\\
    \includegraphics[width=\linewidth]{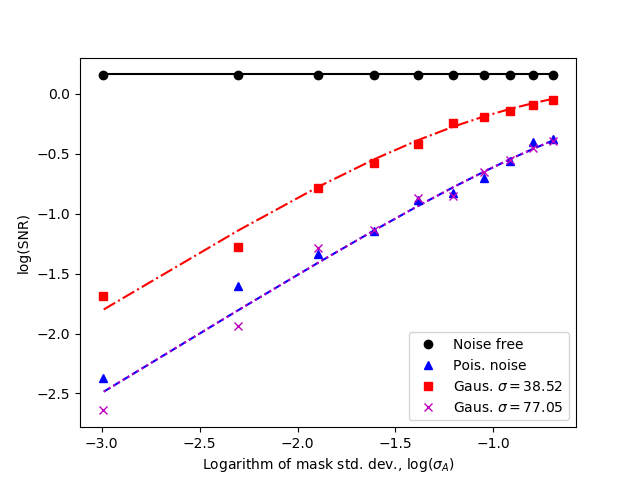}
    \end{minipage}\\
    
    \begin{minipage}{0.5\linewidth}
    \centering
    \scriptsize{(e)}\\
    \includegraphics[width=\linewidth]{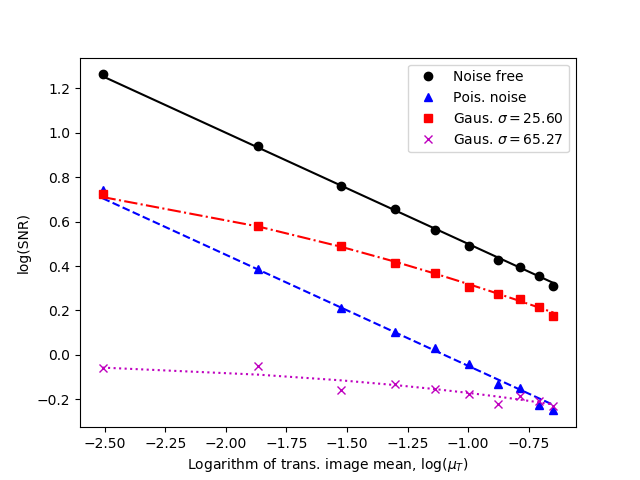}
    \end{minipage}%
    \begin{minipage}{0.5\linewidth}
    \centering
    \scriptsize{(f)}\\
    \includegraphics[width=\linewidth]{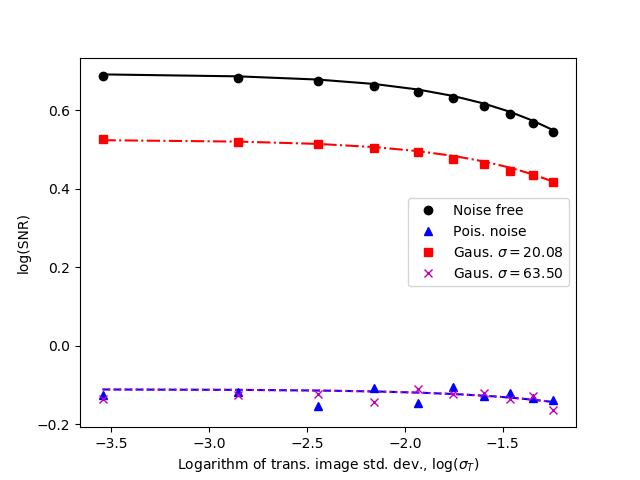}
    \end{minipage}\\
    
    \begin{minipage}{0.5\linewidth}
    \centering
    \scriptsize{(g)}\\
    \includegraphics[width=\linewidth]{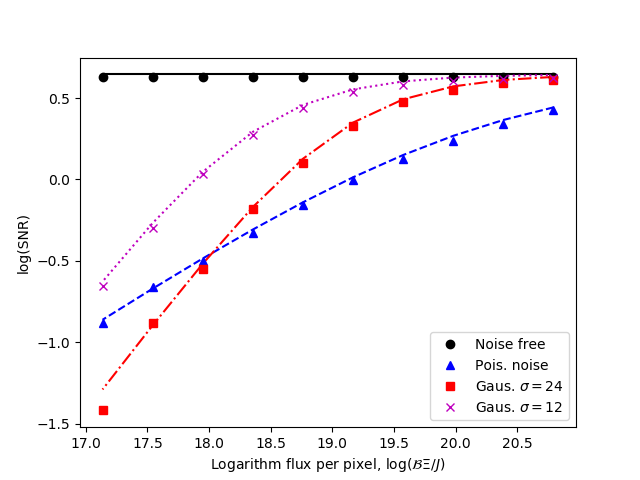}
    \end{minipage}
    
    \caption{Demonstration of the SNR of ghost imaging for random masks under Poisson and Gaussian noise models as a function of imaging parameters: (a) number of masks, $J$; (b) image dimension, $n$; (c) mean mask transmission, $\mu_A$; (d) standard deviation of mask transmission, $\sigma_A$; (e) mean object transmission, $\mu_T$; (e) standard deviation of object transmission, $\sigma_T$; and (g) incident photons per pixel per measurement, $\mathcal{P}/J$. Markers in the plot are simulation results while the lines connect the expected result according to Eq.~(\ref{eq:snrRandom}). Random masks were simulated and the ghost images have been recovered using the scaled adjoint (Eq.~(\ref{eq:adjMatrix})).}
    \label{fig:snrStudyPlots}
\end{figure}

\paragraph{Noisy case with random masks:}  Based on a set of random independent binary masks, Fig.~\ref{fig:snrStudyPlots} shows the response of SNR (Eq.~(\ref{eq:snrRandom})) when varying each parameter while keeping the remainder fixed. Poisson and Gaussian distributions were used to simulate bucket measurements with photon shot-noise and per-measurement electronic noise respectively. For Poisson noise $\sigma_p = 1$, and for the case of Gaussian noise, $\sigma_m$ is set to yield similar noise levels to the Poisson case; three cases of Gaussian noise (one of which is {\it noise free}, i.e., $\sigma_m = 0$) are presented to indicate the trend of relationships.

\begin{figure}[ht!]
    \centering
    \begin{minipage}{0.5\linewidth}
    \centering
    \scriptsize{(a)}\\
    \includegraphics[width=\linewidth]{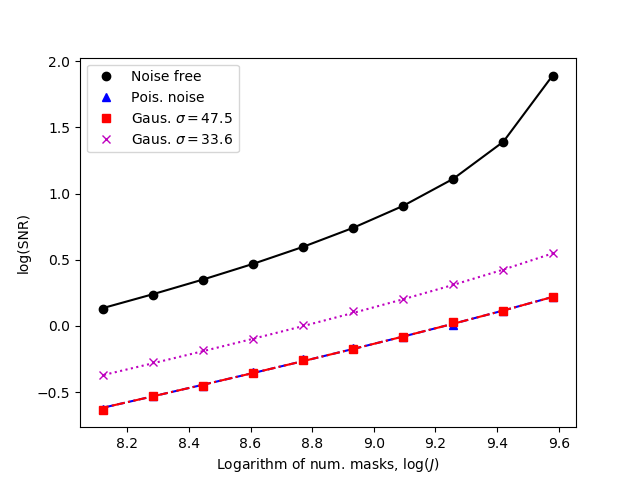}
    \end{minipage}%
    \begin{minipage}{0.5\linewidth}
    \centering
    \scriptsize{(b)}\\
    \includegraphics[width=\linewidth]{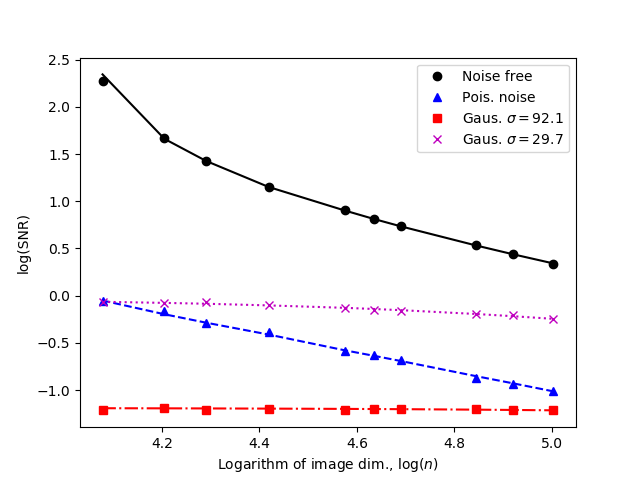}
    \end{minipage}\\
    
    \begin{minipage}{0.5\linewidth}
    \centering
    \scriptsize{(c)}\\
    \includegraphics[width=\linewidth]{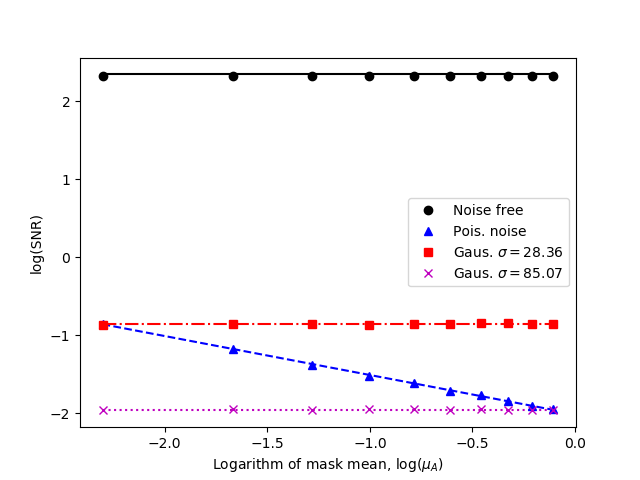}
    \end{minipage}%
    \begin{minipage}{0.5\linewidth}
    \centering
    \scriptsize{(d)}\\
    \includegraphics[width=\linewidth]{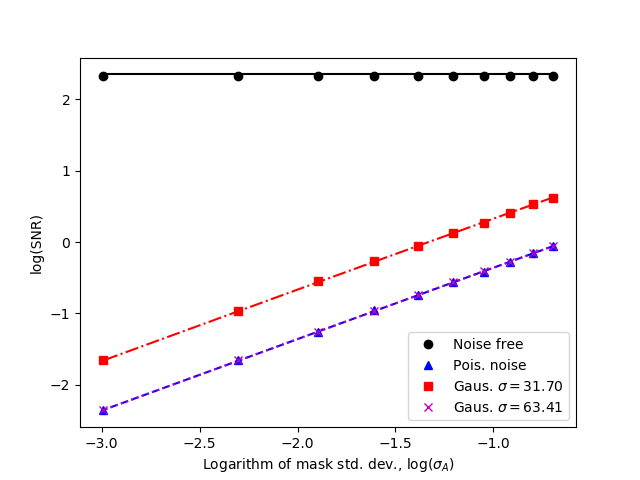}
    \end{minipage}\\
    
    \begin{minipage}{0.5\linewidth}
    \centering
    \scriptsize{(e)}\\
    \includegraphics[width=\linewidth]{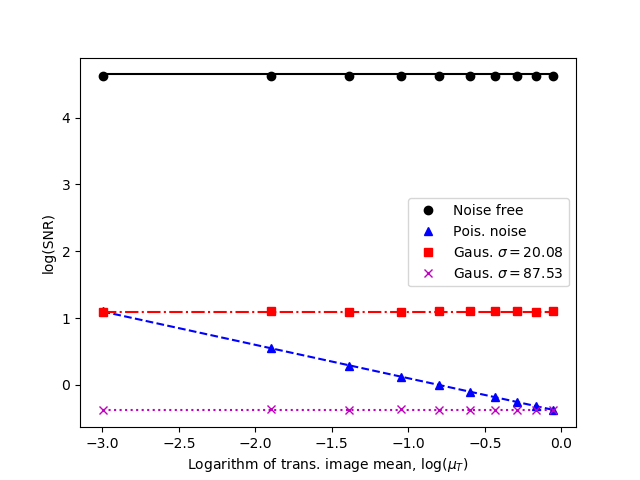}
    \end{minipage}%
    \begin{minipage}{0.5\linewidth}
    \centering
    \scriptsize{(f)}\\
    \includegraphics[width=\linewidth]{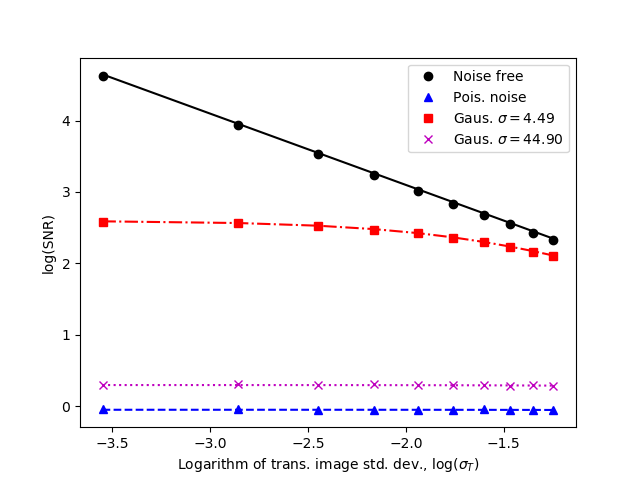}
    \end{minipage}\\
    
    \begin{minipage}{0.5\linewidth}
    \centering
    \scriptsize{(g)}\\
    \includegraphics[width=\linewidth]{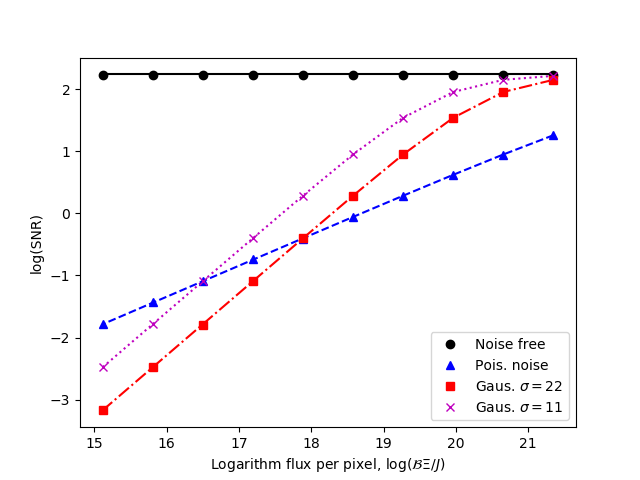}
    \end{minipage}
    \caption{Demonstration of the SNR of ghost imaging for orthogonal masks under Poisson and Gaussian noise models as a function of imaging parameters: (a) number of masks, $J$; (b) image dimension, $n$; (c) mean mask transmission, $\mu_A$; (d) standard deviation of mask transmission, $\sigma_A$; (e) mean object transmission, $\mu_T$; (f) standard deviation of object transmission, $\sigma_T$; and (g) incident photons per pixel per measurement, $\mathcal{P}/J$. Markers in the plot are simulation results while the lines connect the expected result according to the orthogonal variant of Eq.~(\ref{eq:snrRandom}). Finite Radon transform based masks were simulated and the ghost images have been recovered using the scaled adjoint (Eq.~(\ref{eq:adjMatrix})).}
    \label{fig:snrStudyOrthoPlots}
\end{figure}

\paragraph{Noisy case with an orthogonal scanning mask:} Repeating the above simulations but with a set of scanned orthogonal masks, Fig.~\ref{fig:snrStudyOrthoPlots} shows the response of SNR (Eq.~(\ref{eq:snrRandom})) when varying each parameter while keeping the remainder fixed. To prevent $\mbox{SNR} \rightarrow \infty$ the maximum number of masks used was $J_{\mbox{max}} = 0.9 n^2$. Poisson and Gaussian distributions were used to simulate bucket measurements with photon shot-noise and per-measurement electronic noise respectively in a similar manner.

\section{Methods used for simulating the experiments of Zhang {\it et al}.~(2018)}
\label{app:e}

The experiments in Ref.~\cite{zhang2018table} utilized a $5 \times 5$mm$^2$ field-of-view (FOV) with the detector placed 2.5m from the x-ray source. The detector employed for the ultra-low dose experiments was a direct imaging Princeton Instruments PIXIS-XB $1340 \times 1300$ pixel camera with a pixel-pitch of 20$\mu$m. Operating in 2MHz mode, the read-out time, $t_1$, is 0.93s (0.12s when binned $8 \times 8$).

\begin{figure}[ht!]
    \centering
    \begin{minipage}{0.5\linewidth}
    \centering
    \scriptsize{(a)}\\
    \includegraphics[width=0.7\linewidth]{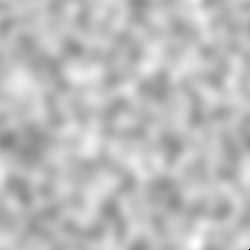}
    \end{minipage}%
    \begin{minipage}{0.5\linewidth}
    \centering
    \scriptsize{(b)}\\
    \includegraphics[width=\linewidth]{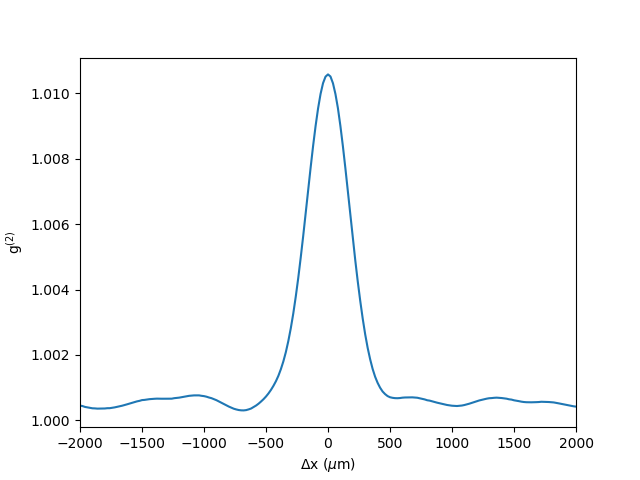}
    \end{minipage}\\
    \begin{minipage}{0.5\linewidth}
    \centering
    \scriptsize{(c)}\\
    \includegraphics[width=0.7\linewidth]{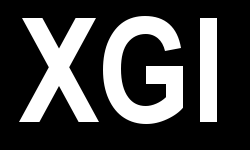}
    \end{minipage}%
    \begin{minipage}{0.5\linewidth}
    \scriptsize{(d)}\\
    \includegraphics[width=0.7\linewidth]{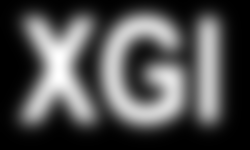}
    \end{minipage}
    \caption{(a) An example speckle image generated as a random binary image with 50\% transmission, blurred by a Gaussian with $\sigma = 0.117$mm. This is then re-scaled to give transmission in the range 50\% - 100\%. The field-of-view (FOV) is $5 \times 5$ mm$^2$. (b) A profile through the point-spread-function (PSF) produced by the set of 10,000 speckle images; x-axis unit is (mm). The PSF has a full-width-at-half-maximum (FWHM) of 0.4mm. (c) The binary transmission image with width 5mm and 20$\mu$m pixel-pitch used for simulation. (d) The expected output transmission image from ghost imaging based on the imaging resolution of 0.4mm.}
    \label{fig:simSpeckle}
\end{figure}

Speckle was generated by sandpaper containing 40$\mu$m diameter silicon-carbide (SiC) grains placed 0.27m from the source. The sandpaper had a minimum transmissivity of approximately 50\% and the grains projected to 0.4mm diameter at the detector. The Gaussian fit to the second-order correlation, $g^{(2)}(x_0; y_0; x; y)$ (defined in Eq.~(2) of \cite{zhang2018table}), was shown to have a maximum of 1.0007 and a full-width-at-half-maximum (FWHM) of 0.4mm. The reference speckle images were recorded with 10s exposure time. We have simulated a set of noise-free speckled illuminations, $A_j(x,y)$ for $j \in [1,J]$, using random binary masks smoothed by a Gaussian function with $\sigma = 0.4/2.355\sqrt{2}=0.12\,$mm. The intensity was scaled to range between 0.5 and 1.0 arbitrary units. An example speckle image is presented in Fig.~\ref{fig:simSpeckle}(a). The second-order correlation generated using 10,000 speckle patterns (Fig.~\ref{fig:simSpeckle}(b)) has a maximum of 1.01 and a FWHM of 0.4mm.

For an object, the experiments in Ref.~\cite{zhang2018table} used a 5mm thick plate of stainless steel with 2.5mm high letters ``CAS'' stenciled out. This has been simulated as a $250 \times 250$px ($5 \times 5$mm$^2$) binary transmission image, $T(x,y)$, with letters ``XGI'' being 100\% transmitting and the remainder being 0\%. This image is presented in Fig.~\ref{fig:simSpeckle}(c) while the expected result (in the noise-free case) from ghost imaging (with 0.4mm resolution) is shown in Fig.~\ref{fig:simSpeckle}(d).

In Ref.~\cite{zhang2018table}, the average flux transmitting through the sandpaper was estimated to be $2.9 \times 10^5$ photons/s/mm$^2$ at the detector.  The flux per pixel was then 120 photons/s and, for an exposure time of $t_0$s, the expected illuminations incident on the CCD were simulated as:
\begin{equation}
\lambda_j(x,y) = T(x,y) \times A_j(x,y) \times 120 t_0,
\end{equation}
for $1 \le j \le J$. The actual image read-out from the CCD was determined as follows:
\begin{eqnarray}
M_j(x,y)
& = & \mbox{photon shot-noise} \nonumber \\
&& \qquad \quad + \mbox{electronic noise}\nonumber \\
& = & \mbox{Poisson}\{\lambda_j(x,y)\}\nonumber \\
&& \qquad \quad + 0.01 \mbox{Poisson}\{100(t_0+t_1)\},
\end{eqnarray}
for $1 \le j \le J$. Here the photon shot-noise is simulated as a Poisson distribution with expected value $\lambda_j(x,y)$; the electronic noise was simulated as a Poisson distribution with an expected signal defined as the equivalent of 10 photons/s per pixel (with 100 electrons per photon) scaled by the exposure time, $t_0$, plus read-out time, $t_1$. For direct (projection) imaging
a read-out time of $t_1 = 0.93$s was used, however, while bucket measurements were made the CCD was simulated to be configured for $8 \times 8$ binning with a read-out time of $t_1=0.12$s.

\begin{figure}[ht!]
    \centering
    \scriptsize{(a)}\\
    \includegraphics[width=0.8\linewidth,angle=180,origin=c]{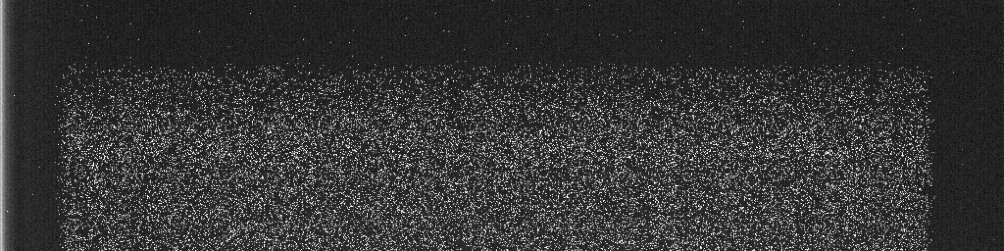}\\
    \scriptsize{(b)}\\
    \includegraphics[width=0.8\linewidth,angle=180,origin=c]{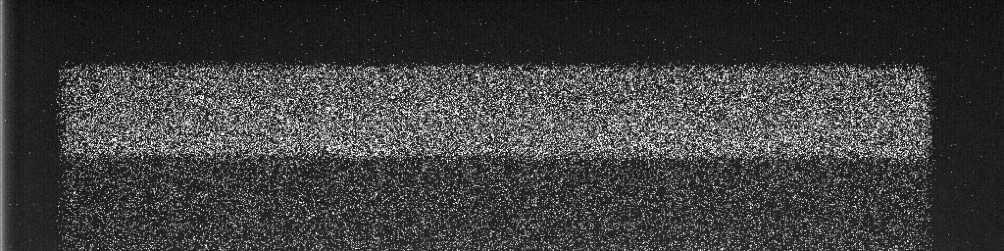}\\
    \scriptsize{(c)}\\
    \includegraphics[width=0.8\linewidth,angle=180,origin=c]{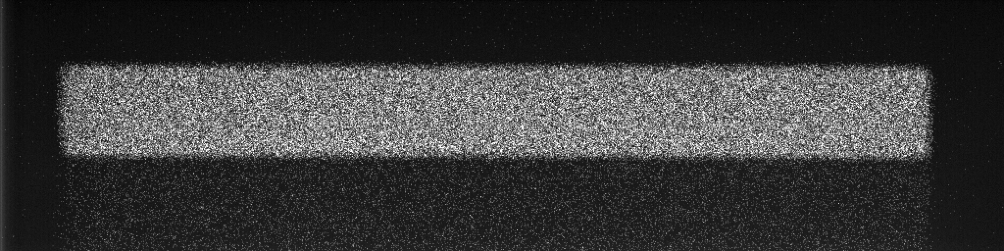}
    \caption{Average of 20 images of x-ray transmission through a slot in a steel computer grill resulting from CCD read-out with a CW x-ray source and no shutter using exposure times of (a) 0.01s, (b) 0.1s, and (c) 0.4s. The CCD read-out time is 0.111s. Slot dimension is 20mm $\times$ 2mm.}
    \label{fig:wilfExpNoShutter}
\end{figure}

\section{CCD read-out process under the presence of a CW x-ray source}
\label{app:f}

Measurement using a {\it charge-coupled device} (CCD) is typically divided into an exposure phase, followed by a read-out phase. During exposure: the CCD passively converts incoming photons to electrons which are stored in its cells. During read-out: the total charge stored in the cells is recorded one line at a time. After each line of cells is read-out, the entire array of cells is shifted down one line and the next line is read-out; this process is repeated until the cells in all lines have been read. In the presence of a continuous wave (CW) x-ray source, if no shutter is employed, cells continue to convert incoming photons during this cell-shifting process to additional electrons building up charge that is later converted to unwanted signal. These errors are commonly referred to as ``vertical smear'' since the additional charge accumulates in vertical columns of cells (matching the cell shifting process).

Here we present a straightforward experiment to illustrate the ``read-out smear'' when using a CW x-ray source on an x-ray CCD camera. Here we used an Andor Newton camera with an e2v CCD30-11 front-illuminated deep-depletion CCD sensor that has a $1024 \times 256$ pixel array and a 26$\mu$m pixel pitch. Figure \ref{fig:wilfExpNoShutter} presents the results of imaging the transmission through a slot in a steel computer grill. Adjacent slits are blocked using 2mm thick Pb. The source-detector distance is 100mm. Three exposures of increasing time have been presented: 0.01s, 0.1s, and 0.4s. The time between adjacent frames (``kinetic cycle time'') corresponding to each of these measurements is 0.121s, 0.211s, and 0.511s, due to a non-zero read-out time of $t_1 = 0.111$s. In each case the images presented in Fig.~\ref{fig:wilfExpNoShutter} are the average over 20 consecutive frames.

Looking at the 0.01s exposure in Fig.~\ref{fig:wilfExpNoShutter}(a), catching this 10ms exposure window is trivial using a pulsed source, but with a CW source and read-out time more than 10 times longer than that window, we observe a uniform ``smear'' due to the CCD reading out the image line-by-line and shifting the image down. For the 0.1s exposure in Fig.~\ref{fig:wilfExpNoShutter}(b), the exposure becomes significant in terms of the read-out time, so the relative significance of the smear begins to fade. The smear becomes less significant as exposure time is increased beyond this; the smear has become difficult to see in Fig.~\ref{fig:wilfExpNoShutter}(c) with 0.4s exposure time.

\end{document}